\documentclass[prd,aps,nofootinbib]{revtex4}
\usepackage{slashed}
\usepackage{graphicx}
\usepackage{amsmath}
\usepackage{amssymb}
\usepackage{amsfonts}
\usepackage{indentfirst}
\usepackage{color}

\newcommand{\be}{\begin{equation}}
\newcommand{\ee}{\end{equation}}
\newcommand{\bea}{\begin{eqnarray}}
\newcommand{\eea}{\end{eqnarray}}

\newcommand{\f}{\frac}
\begin{document}

\preprint[\leftline{KCL-PH-TH/2023-{\bf 56}}

\title{\Large {\bf Effective Abelian Lattice Gauge Field Theories for scalar-matter-monopole interactions}}

\vspace{0.5cm}

\author{K. Farakos$^{a}$, G. Koutsoumbas$^a$ and Nick E. Mavromatos$^{a,b}$}, 

\vspace{0.5cm}

\affiliation{$^a$National Technical University of Athens, 
School of Applied Mathematical and Physical Sciences, Department of Physics, Zografou Campus GR157 80, Athens, Greece}

\vspace{0.5cm}

\affiliation{$^b$ Theoretical Particle Physics and Cosmology Group, Physics Department, King's College London, Strand, London WC2R 2LS, UK.}

\begin{abstract}

\vspace{0.5cm}

We present a gauge and Lorentz invariant effective field theory model for the interaction of a  charged scalar matter field with a magnetic monopole source, described by an external magnetic current. The quantum fluctuations of the monopole field are described effectively by a strongly-coupled ``dual'' $U_{\rm d}(1)$ gauge field, which is independent of the electromagnetic $U_{\rm em}(1)$ gauge field. The effective interactions of the charged matter with the monopole source are described by a gauge invariant mixed Chern-Simons-like (Pontryagin-density) term between the two $U(1)$ gauge fields. The latter interaction coupling is left free, and a Lattice study of the system is performed with the aim of determining the phase structure of this effective theory. Our study shows that, in the spontaneously-broken-symmetry phase, the monopole source triggers, via the mixed Chern-Simons term, which is non-trivial in its presence, the generation of a dynamical singular   configuration (magnetic-monopole-like) for the respective gauge fields. The scalar field also behaves in the broken phase in a way similar to that of the scalar sector of the `t Hooft-Polyakov monopole.

\end{abstract}

\maketitle

\section{Introduction and Motivation \label{sec:intro}} 

The structureless magnetic monopole of 
Dirac~\cite{Dirac}, was characterised by the presence of the ``Dirac string'', which furnishes the theory with Lorentz-violating non-local hidden degrees of freedom. The Dirac 
 charge quantization condition leads to the invisibility of the Dirac string. 
 The Lorentz violation necessitated by the presence of the string is manifested in various field theoretic concepts contexts, such as  the local formulation of the magnetic charges by Zwanziger~\cite{Zwanziger:1970hk}, which avoids the initial non-local degrees of freedom of the Dirac string by the presence of a fixed four vector in the associated effective Lagrangian, which involves two gauge potentials, associated with electric and magnetic current sources, or Weinberg's paradox~\cite{Weinberg}, which states that the leading perturbative term in the scattering amplitude between a magnetic pole and an electric charge has a non-Lorentz invariant form.\footnote{For earlier attempts to discuss quantum electrodynamics in the presence of magnetic monopoles see \cite{cab}. That work also makes use of two gauge potentials as in \cite{Zwanziger:1970hk}, but  the pathologies associated with the Dirac string are avoided using path-dependent variables associated with the field strengths rather than the gauge potentials. In our approach, using the formalism developed initially by \cite{Zwanziger:1970hk}, we also avoid the Dirac string, as we discussed in \cite{Alexandre} (using essentially the argumentation of \cite{terning}), and shall review below.}  
After Dirac, Schwinger'~\cite{schw} has  generalised the magnetic monopole to objects, called dyons, which contain both electric ($q_e$) and magnetic ($q_m$) charges, restoring Lorentz invariance, but at the unavoidable introduction of a non-local Hamiltonian. The restoration of Lorentz symmetry is guaranteed upon the imposition of Schwinger generalisation of Dirac's 
quantization condition in the scattering of two dyon configurations (1,2) with electric ($q_e^{(i)}$) and magnetic charges ($q_m^{(i)}$), where $i =1,2$ labels the dyons (with the convention, positive charges for particles, and negative for antiparticles - in this article we follow the notation of \cite{Zwanziger:1970hk}):
\begin{equation}\label{schquant}
\Big(q^{(1)}_e\, q^{(2)}_m  - q^{(2)}_e \,q^{(1)}_m \Big)/4\pi \, = \, {\mathcal Z}_{em} \in \mathbb{Z}~, 
\end{equation}
where ${\mathbb Z}$ is the set of integers. 

In this article we shall restrict ourselves  to the case of a magnetic monopoles, which is obtained from \eqref{schquant} in the particular case 
of, say, a particle (1) corresponding to an ordinary electrically-charged matter particle, carrying only  electric charge, $q^{(1)}_e = q_e$, while particle (2) is a magnetic monopole, carrying only magnetic charge  $q^{(2)}_m = q_m$ . Then, 
the condition \eqref{schquant} reduces to the standard Dirac quantisation,
\begin{equation}\label{dirac}
\Big(q_e \, q_m  \Big)/4\pi \, = \,  \, {\mathcal Z}_{em} \in \mathbb{Z}~.
\end{equation}
The reader should notice that the fundamental charge unit is twice that of Dirac~\cite{Dirac}, given that in the Dirac case the right-hand side of \eqref{dirac} would be ${\mathcal Z}/2$.
Indeed, for the case where $q_e=e$, the electron charge, the magnetic charge assumes the 
value 
\begin{align}\label{magch}
q_m=  \frac{4\pi}{e^2} e \,  {\mathcal Z}_{em} =  \frac{1}{\alpha} e\,  {\mathcal Z}_{em} \equiv 2\, {\mathcal Z}_{em} \, g_D \,, \qquad  {\mathcal Z}_{em} \in \mathbb Z\,, 
\end{align}
where $g_D= \frac{1}{2\alpha} \, e = 68.5 \, e$ is the fundamental unit of magnetic (Dirac) charge, with $\alpha=1/137$ the fine structure constant (at zero energy scale). 

It should be remarked that, upon the imposition of \eqref{schquant} or \eqref{dirac},  any Lorentz non-invariant effect in the effective two-gauge-potential Lagrangian of 
Zwanziger~\cite{Zwanziger:1970hk} disappears. This feature is also associated with an integrability condition of the representation of the Poincar\'e Lie algebra, that stems from Poincar\'e invariance into a representation of the finite Poincar\'e group~\cite{Zwanziger:1970hk}.

Long after Dirac's proposition of structureless monopoles, `t Hooft and Polyakov~\cite{thooft}
proposed {\it composite} monopoles, which were (finite-energy) {\it topological-soliton} solutions 
of the pertinent Euler-Lagrange equations of motion of 
gauge and Lorentz invariant field theories, with spontaneous (Higgs-like) symmetry breaking.  It is important to remark that such composite monopoles satisfied the condition \eqref{dirac} but with the fundamental charge unit being twice as that of Dirac. Unlike the Dirac case, they are smooth field configurations which do not have Dirac strings. 
The solution is localised around the origin (monopole center), where the gauge group is unbroken. On the other hand, asymptotically far from the center of the monopole the gauge group G breaks spontaneously to a subgroup H.
At such large distances 
the `t Hooft-Polyakov monopole behaves as a Dirac one.  In his construction of the SU(2) monopole, `t Hooft considered the Georgi-Glashow model~\cite{gg} involving a  Higgs triplet that spontaneously breaks the SU(2) group. The simply connected SU(2) gauge group is exhibits a non-trivial homotopy, $\pi_2$(SU(2))$= {\mathbb Z}$, with ${\mathbb Z}$ the set of integers, defining the number of times the spatial three-sphere which the monopole and its constituent fields  
live on,  wraps around the internal(gauge)-space sphere spanned by the Higgs triplet of the Georgi-Glashow model. This leads to the magnetic charge quantization condition (\eqref{schquant} 
for the magnetic monopole. Monopole and dyon solutions of 
 phenomenologically realistic Grand Unified Theories (GUT) with gauge group SU(5), having large masses of order of the GUT scale $10^{14}-10^{16}$ GeV, were discussed in \cite{dokos}. Inflation of the Universe at such scales, should wash out these heavy monopoles thus
 providing a natural explanation of their absence from the Cosmos today, consistent with the null results of the pertinent 
cosmic searches so far~\cite{patrizii}.  Magnetic monopoles exist also in superstring theories~\cite{wen}, as well as in D-brane-inspired GUT models~\cite{shafi}. In the latter models, the unification scale, and therefore the magnetic monopole/dyon mass, 
can be lowered significantly down to $10^4-10^6$ GeV,  which might be relevant for future collider or cosmic-ray searches of such objects. 
For other discussion involving topological structures in beyond-the-standard-model theories, the reader is referred to the recent literature~\cite{Lazarides}.

Unfortunately, unlike the SU(2) or SU(5) or other GUT-like-group monopoles, the gauge group of the standard model (SM) SU(2)$\times$ U$_Y$(1), does not have this  simple structure due to the hypercharge U$_Y$(1) factor. As a result, after Higgs breaking, the quotient group SU(2)$\times$ U$_Y$(1)/U$_{\rm em}$(1) is not characterised by a non-trivial second homotopy, thus monopoles were not expected to exist in the Standard Model. 
However, in \cite{cho}, it was argued that  one can look for non-trivial homotopy features not in the gauge but in the Higgs field sector of the model. Indeed, in this case the Glashow-Weinberg-Salam model with a Higgs sector is viewed as a gauge $CP^1$ model with the (normalised) Higgs doublet field playing the role  of the corresponding CP$^1$ field. The latter is characterised by a non-trivial homotopy $\pi_2(CP_1)={\mathbb Z}$, thus allowing in principle for a topological quantization \`a la `t Hooft-Polyakov, and thus the existence of magnetic monopoles/dyons. However, the resulting monopole or dyon solutions of \cite{cho} have infinite energy.
Finite energy monopoles of the type proposed in \cite{cho} can characterise extensions of the Standard Model, with either appropriate non-minimally coupled Higgs and hypercharge sectors~\cite{emy}, or higher-derivative extensions of the hypercharge sector, for instance a (string-inspired) Born-Infeld configuration~\cite{aruna}. Such monopole/dyon  solutions  could have masses accessible to the scales of current or future colliders. Other finite-energy structured monopole/dyon solutions with potentially low mass can be found in string-inspired models with axion-like structures~\cite{sarkar}, or models of neutrino masses~\cite{hung}, beyond the standard model of particle physics, with non-sterile right-handed,  whose electroweak-scale Majorana masses are obtained by the coupling to a complex Higgs-like triplet of scalar fields. A recent review of such magnetic monopole solutions, and their experimental searches in colliders and in the Cosmos,  is given in \cite{mmmono}. 
Moreover, there have been interesting recent studies~\cite{Lazarides2} in discussing novel magnetic monopole-like structures (with sufficiently low masses) upon embedding the Standard Model into GUT models, which consist of an appropriate merging between Dirac-like structures with Nambu monopoles~\cite{nambu}, the latter being string-like structures of the electroweak theory, which utilize a Higgs doublet field in the SU(2) sector that behaves like a scalar triplet under SU(2) with zero hypercharge.\footnote{We note, for completion, that,  in the theory of \cite{nambu}, the existence of the monopole solutions leads to the so-called Nambu dumb-bell configurations of monopole/antimonopole pairs connected via a Z-flux string. In the constructions of \cite{Lazarides2}, one can have composite structures consisting of Dirac monopoles connected to one or several Nambu-monopoles via Z-flux strings. To ensure well defined (finite) energies, though, of such configurations, one needs, as in the case of \cite{cho} mentioned above, to embed the electroweak theory into appropriate extensions of the standard model, such as GUT.}

In view of the above theoretical evidence for the existence of of relatively light monopoles/dyons, there emerges a pressing need for the development  of appropriate effective field theory models and methods 
that could allow for the study of the production of such objects at colliders or their scattering off standard model matter, like quarks and leptons. Lacking at present a fundamental theory for the description of such interactions, one can employ {\it ad hoc} phenomenological effective U(1) gauge field theory models, usually based on appropriate duality symmetries~\cite{monprod}.  
Indeed, such U(1) models are essentially dual descriptions of electrically charged particles of various spins interacting with photons, in which the electric charge that appears in the interaction with photons is replaced by an effective magnetic charge. As discussed in \cite{drukier}, one may think of the magnetic monopole charge in such cases as a {\it collective coupling} to photons of (electrically) charged constituent degrees of freedom, such as charged $W$-bosons and Higgs fields, which the monopole is composed of. Modeling these constituent fields as quantum harmonic oscillators, the authors of \cite{drukier} argued that the 
monopole might be viewed as a {\it coherent superposition} of 
$\sim \frac{1}{\alpha}$ such quantum states, with the result that the collective coupling to photons is $\frac{1}{\alpha} \, e$, quantization condition \eqref{schquant}. Such a representation also leads to a significant suppression of the production cross section of such composite magnetic monopoles in colliders. This conclusion does not apply to the case of structureless Dirac monopole sources.
At this point it worths mentioning that such suppression is avoided in 
the Schwinger production of magnetic monopole/antimonopole pairs (with or without structure)~\cite{schwing}, which is a non-perturbative mechanism, thereby providing reliable mass bounds for monopole masses in interpretation of experimental searches for such objects at colliders~\cite{schwingmoedal}.

In \cite{Alexandre} a first attempt was made to construct (strongly coupled) effective gauge field theories for the abovementioned type of composite monopoles, by extending non trivially the ideas of 
Zwanziger~\cite{Zwanziger:1970hk} that were developed for structureless monopoles, using a novel Schwinger-Dyson renormalisation treatment. The effective quantum field theory invoked two $U(1)$ gauge fields, one $U_{\rm em}(1)$ characterised by a weak coupling, which plays the r\^ole of electromagnetic interactions, and the other, a (strongly) coupled  ``dual'' gauge $U_d(1)$ field, which expresses the quantum fluctuation of the classical dual potential of \cite{Zwanziger:1970hk}. In \cite{Alexandre} we used the quantum $U(1)$ gauge fields as {\it independent} path integration variables.

When the effective theory is considered sufficiently far away from the monopole centre, the `t-Hooft-Polyakov-type monopoles mentioned above resemble the structureless Dirac ones, to a good approximation. Nonetheless, as explained in \cite{Alexandre}, the selected Schwinger-Dyson approach implied that the method was appropriate {\it only} for 
composite monopoles, since when one considers the  quantisation of our non-perturbative effective field theory, the resulting wave-function renormalization  for the monopole field will not respect the appropriate unitarity bounds for an elementary field, thus making our effective description suitable only for composite fields~\cite{psbook}. The monopoles of the work of \cite{Alexandre} were assumed, for definiteness, to be fermionic. 

In this work we follow another effective approach of a theory with two $U(1)$, which however concentrates only  on the effects of the interaction of matter with a background of a monopole source, with the quantum fluctuations of the latter being represented by  a (strongly coupled in general) ``dual'' $U_d(1)$, which is independent of the electromagnetic interactions, as in the model of \cite{Alexandre}, the latter interacting only with the charged matter field, assumed to be a scalar field. 
However, we introduce an interaction, of Chern-Simons(CS) (Pontryagin-density) type , between the two $U(1)$'s, whose origin is inspired by, but is actually a generalisation of, the constraint of  the model of  \cite{Zwanziger:1970hk} among the  {\it classical} gauge potentials of its two $U(1)$'s, so that in the classical limit of the theory there is only one degree of freedom, that of the photon.

Concerning the nature of the monopole, we shall be quite generic in our considerations, by not specifying whether it is composite or point-like. For us, the monopole might be one of the known microscopic types, mentioned above, or an as yet unknown solution of some beyond the SM theory with or without Dirac string singularities. Its spin is also not going to be specified. The resummation of the strongly-coupled dual $U_d(1)$ sector of the theory  in our case below will be done by placing the theory on a {\it Lattice}.

The structure of the article is as follows: in the next section \ref{sec:model} we present the continuum model for the effective description of the dynamics of a quantum-fluctuating monopole interacting with scalar Higgs matter, and set up the formalism, which is based on a $U_{\rm em}(1) \times U_d(1)$ gauge theory, with a scalar sector of Higgs type, associated with the electromagnetic $U_{\rm em}(1)$ only. In section \ref{sec:lattice} we discuss the lattice action corresponding to the continuum theory of section \ref{sec:model}. In section \ref{sec:higgs} we discuss spontaneous symmetry breaking in the Higgs sector and the r\^ole of the monopole background. We discuss the symmetric and spontaneously broken phases of the theory, as well as the dependence of the results on the strength of the characteristic coupling parameter $\xi$ of the model introduced by the appropriate constraint between the gauge potentials of the two $U(1)$'s in the model. The dependence of the effects of the monopole on the various fields configurations in regions near and away of the monopole core is discussed in subsection. For completeness, and to stress the effects of the characteristic kinetic mixing between the field strength of the electromagnetic $U_{\rm em}(1)$ and the {\it dual field strength} of the $U_d(1)$, which is crucial for the model, and is induced by an appropriate constraint responsible for the appearance of the coupling $\xi$, we also consider in section \ref{sec:FAFC} a model with a normal kinetic mixing between the two $U(1)$'s instead of the aforementioned axial kinetic mixing. We repeat the analysis of section \ref{sec:higgs} for this case in section \ref{sec:FAFC}, including again a study of the corresponding  broken phase, and a discussion on the $\xi$-dependence of the corresponding results. A comparison of the results between sections \ref{sec:higgs} and  \ref{sec:FAFC} follows, where the non-trivial effects of the magnetic monopole background on the phase diagram of the model of sections \ref{sec:model} and \ref{sec:lattice} are distinguished from the case of section \ref{sec:FAFC}, which involves ordinary kinetic mixing between the two $U(1)$'s. Finally,  conclusions and outlook are presented in section \ref{sec:concl}. In the appendix, section \ref{sec:simul}, we discuss the details of our lattice simulations.

\section{The Effective Field Theory modelling scalar-matter-monopole interactions}\label{sec:model}

Our study will be based partly on the work of Zwanziger~\cite{Zwanziger:1970hk}, which we shall review below for completeness. 
The approach employs two related gauge fields, whose existence avoids the use of non-local Dirac strings, but at the cost of having 
Lorentz-violating terms in the pertinent local Lagrangian describing the dynamics of monopoles/dyons.
If one considers electric and magnetic currents, $J_e^\mu$ and $J_m^\mu$ respectively,  
 then, as shown in \cite{Zwanziger:1970hk}, the corresponding Maxwell's equations read
 \begin{equation}\label{maxwell}
 \partial_\mu \, F^{\mu\nu} = J_e^\nu, \quad \partial_\mu \, ^\star F^{\mu\nu} = J_m^\nu, 
 \end{equation}
where $F^{\mu\nu}$  is the electromagnetic field-strength tensor, while $^\star F^{\mu\nu} \equiv \frac{1}{2}\epsilon^{\mu\nu}_{\,\,\,\,\,\,\rho\sigma}\, F^{\rho\sigma}$ is the dual tensor; 
$\epsilon^{\mu\nu\rho\sigma}$ denotes the totally antisymmetric Levi-Civita symbol, with $\epsilon^{0123}=+1$, etc. Throughout this article, we work
 in a flat Minkowski 
space-time with metric $\eta^{\mu\nu}=(1, -1, -1, -1)$. For future use, the reader should keep a note of the the {\it axial}-vector (pseudovector) nature of the magnetic current in \eqref{maxwell}.

\subsection{The two-potential formalism}\label{sec:2potform}

The general solution of the Maxwell's equations \eqref{maxwell}, is expressed in terms of two 
{\it classical} potentials $\mathcal A_\mu$ and $\mathcal C_\mu$
and a fixed four vector $\eta^\mu$, as follows~\cite{Zwanziger:1970hk}:
\begin{equation}\label{gensol} {\rm First~eq.~with~electric~current:} \, F = - ^\star (\partial \wedge \mathcal C) + (\eta \cdot \partial)^{-1} \, 
(\eta \wedge J_e), \quad ^\star F= \partial \wedge \mathcal C + (\eta \cdot \partial)^{-1} \, ^\star (\eta \wedge J_e), 
\end{equation}
\begin{equation}\label{gensol2}
{\rm Second~eq.~with~magnetic~current:} \, ^\star F = ^\star (\partial \wedge \mathcal A) + (\eta \cdot \partial)^{-1} \, (\eta \wedge J_m), 
\quad F= \partial \wedge \mathcal A - (\eta \cdot \partial)^{-1} \, ^\star (\eta \wedge J_m),
\end{equation}
where we used a differential form notation for brevity, in which $\wedge$ ($\cdot$) denotes exterior (interior) product, whose action on  four vectors is defined as:
$(a \wedge b)^{\mu\nu} \equiv a^\mu b^\nu - a^\nu b^\mu$, $a \cdot b \equiv a^\mu b_\mu$.   With these conventions we have $\mathcal F_{\mu\nu}=- \mathcal F_{\nu\mu} $: $^\star \,^\star \mathcal F_{\mu\nu} = - \mathcal F _{\mu\nu}$,  for any antisymmetric second-rank tensor. 

We note that the currents can be eliminated from \eqref{gensol} and \eqref{gensol2}~\cite{Zwanziger:1970hk}, 
so that these equations can be expressed only in terms of the classical potentials $\mathcal A_\mu$ and $\mathcal C_\mu$. One also uses 
the following representation of the kernel $(\eta \cdot \partial)^{-1}(x)$ (satisfying 
$\eta \cdot \partial \, (\eta \cdot \partial)^{-1}(x) = \delta^{(4)}(x)$):
\begin{equation}\label{kernel}
(\eta \cdot \partial)^{-1}(x) = c_1 \int_0^\infty \delta^{(4)}(x - \eta \, s) \, ds - (1 - c_1) \,  \int_0^\infty \delta^{(4)}(x +\eta \, s) \, ds,
\end{equation}
with $c_1$ a real constant,  appropriately defined in order to obtain the correct form of the Lorentz force  in the classical relativistic particle limit of the dyon field~\cite{Zwanziger:1970hk}. 
The form \eqref{kernel} implies that, in the point-particle case, the support of $(\eta \cdot \partial)^{-1}(x_i - x_f)$ is reduced to $x_i^\mu (\tau_i) -  x_f^\mu (\tau_f) = \eta^\mu s$, 
for $-\infty < \tau_i, \tau_f, s < +\infty$, with $\tau$ the proper time. 

The classical gauge potentials $\mathcal A_\mu$ and $\mathcal C_\mu$ depend on $\eta^\mu$ and on the gauge choice. For convenience, in the approach of ref.~\cite{Zwanziger:1970hk}, 
the fixed four vector $\eta^\mu$ was chosen to be space-like $\eta^\mu \eta_\mu < 0$. The gauge potentials are not independent, being associated with a single field strength $F$, 
since $^\star F$ is expressed in terms of $F$. 
Indeed, from \eqref{gensol} and \eqref{gensol2} one can see that~\cite{Zwanziger:1970hk}:
\begin{equation}\label{relAB}
\partial \wedge \mathcal A + ^\star (\partial \wedge \mathcal C) = (\eta \cdot \partial)^{-1} \Big[ \eta \wedge J_e + ^\star (\eta \wedge J_m)\Big],
\end{equation} 
 so that the two gauge potentials are related, which as already mentioned, leads to the fact that the classical local theory of \cite{Zwanziger:1970hk} has only two on-shell photon propagating degrees of freedom. From the expression \eqref{kernel}, it becomes clear that the gauge potentials assume non trivial values along the direction of the Dirac string $\eta^\mu$. 
 The constraint \eqref{relAB} would imply that $\mathcal C$ is an {\it axial vector}, since this is the case with the magnetic current as well, and in this way one obtains consistent 
 transformations under the (improper) Lorentz group, which includes spatial reflections. This will be crucial for our purposes in the next section, where we develop the effective 
 quantum gauge field theory for the interaction of a magnetic monopole with charged scalar matter. 
 
 We mention at this point that the corresponding Lagrangian proposed in \cite{Zwanziger:1970hk} describing the classical
 dynamics of magnetic monopoles, in the presence of Lorentz- and gauge-invariance violating  $\eta^\mu$-dependent terms, is given by:
 \begin{eqnarray}\label{lagZ}
{\mathcal L} &=& \frac{1}{8}{\rm tr}\Big[(\partial \wedge \mathcal A) \cdot (\partial \wedge \mathcal A)\Big] +  
\frac{1}{8}{\rm tr}\Big[(\partial \wedge \mathcal C) \cdot (\partial \wedge \mathcal C)\Big] - J_e \cdot \mathcal A - J_m \cdot \mathcal C \nonumber \\
&-& \frac{1}{4\eta^2}\, \Big(\eta \cdot \Big[(\partial \wedge \mathcal A)  + ^\star (\partial \wedge \mathcal C)\Big]\Big)^2  - 
\frac{1}{4\eta^2}\, \Big(\eta \cdot \Big[(\partial \wedge \mathcal C)  - ^\star (\partial \wedge \mathcal A)\Big]\Big)^2~.
\end{eqnarray}
where we used the notation $${\rm tr}({\mathcal F} \cdot {\mathcal F}) \equiv {\mathcal F}_{\mu\nu} {\mathcal F}^{\nu\mu}$$ for any second rank tensor $\mathcal F$. 

The presence of the monopole, and its topologically non-trivial nature (that is solutions of a certain field theory in non-trivial
sectors with monopole number $n =1,2, \dots$) is reflected precisely on the impossibility to deform continuously the vector 
so as $\eta^\mu \to 0^\mu \equiv (0,0,0,0)^T$, with $T$ denoting matrix transposition. 
However, we note that, on using the form \eqref{kernel}, such a limit can be {\it formally} taken in \eqref{relAB}, \eqref{lagZ}, 
since we can formally set the $\eta$-dependent right-hand-side to zero (given that it approaches $0$ faster than $\eta^2 \to 0$, 
given that the operator $(\eta \cdot \partial)^{-1}(x_i - x_f)$ has zero support in the limit $\eta^\mu \to 0^\mu$). In such a limit, 
\eqref{relAB} leads to:
\be\label{solconstr}
\partial \wedge \mathcal A  + ^\star (\partial \wedge \mathcal C)\stackrel{\eta^\mu \to 0^\mu}{=} 0~, 
\ee
The Lagrangian (\ref{lagZ}) then does not contain any 
Lorentz-symmetry violating term, and involves two gauge fields related by the constraint (\ref{solconstr}).  
 
One could physically justify the limit \eqref{solconstr}, by recalling the study of ref.~\cite{terning}. There it was argued that the Lorentz-violating effects of a magnetic pole associated with 
the Dirac string can be resummed in a non-perturbative way with the result that 
the scattering amplitude of an electrically-charge matter particle off a magnetic charge contains all such Lorentz-violating effects in a phase.
The latter thus drops out of physical quantities such as cross sections. Additionally, such a phase turns out to be a multiple of $2\pi$ if  the quantization condition \eqref{schquant} is valid, in which case  the corresponding amplitude is Lorentz invariant. Although 
such a conclusion was reached by means of studying a toy model employing perturbative magnetic charges in a dark sector, using the two potential formalism \eqref{lagZ} appropriately 
in both the visible and dark sectors, and assuming a perturbative small mixing of ordinary photons with dark photons, nonetheless we shall assume for our purposes in this work, following \cite{Alexandre}, that a similar conclusion characterises the realistic non-perturbative magnetic monopole cases. This in turn leads to a decoupling of the 
Lorentz-violating-(Dirac-string-like) effects  of the vector $\eta^\mu$ from the relevant cross sections. 

Hence, as in \cite{Alexandre}, from now on we shall ignore $\eta$-dependent terms in the pertinent effective Lagrangian, but assume the constrain \eqref{solconstr} in the pertinent effective quantum field theory, which we next proceed to develop. We stress once again that the effective field theory model to be discussed below contains only the magnetic monopole as a source, without specifying its microscopic structure and spin content, or the underlying field theory which admits is as a solution of the relevant equations of motion. If our model turns out to be correct, then all the above-described models can be used as provider of such source terms.

\subsection{The Continuum Model}\label{sec:contmodel}
 
The constraint \eqref{solconstr} is now a gauge-invariant one, and can be implemented in the corresponding {\it Euclidean} path integral, involving charged matter fields coupled directly only to the electromagnetic potential $\mathcal A$,  
by means of a $\delta$-functional constraint of the Lorentz-invariant square of the left-hand-side of \eqref{solconstr}. The corresponding effective action is given by \eqref{lagZ}, in the limit $\eta^\mu \to 0^\mu$, so the relevant partition function reads:
{\small \begin{align}\label{cPI}
\mathcal Z &=  \int D {\mathcal A} \, D {\mathcal C} \exp\Big( i\, \int d^4 x (-\frac{1}{4} F_{\mu\nu}(\mathcal A)\, F^{\mu\nu}(\mathcal A) -\frac{1}{4} \mathcal G_{\mu\nu}({\mathcal C})\, \mathcal G^{\mu\nu}({\mathcal C})) + \mathcal L_{\rm matter}(\phi^\dagger, \phi, \mathcal A) 
+ \dots 
\Big) \, \delta\Big[ (F_{\alpha\beta} + \widetilde{\mathcal{G}}_{\alpha\beta})\,(F^{\alpha\beta} + \widetilde{\mathcal{G}}^{\alpha\beta})\Big]\,,
\end{align}}where $F_{\mu\nu} = \partial_\mu \mathcal A_\nu - \partial_\nu \mathcal A_\mu $ is the Maxwell tensor of $U_{\rm em}(1)$, 
$\widetilde{\mathcal G}_{\mu\nu} = \frac{1}{2} \epsilon_{\mu\nu\rho\sigma} \, \mathcal{G}^{\rho\sigma}$ denotes the dual of $\mathcal{G}_{\mu\nu}
 = \partial_\mu \mathcal C_\nu - \partial_\nu \mathcal C_\mu$,\footnote{The reader should recall that, in differential form notation, we have denoted previously the dual of $ \mathcal G$ by a Hodge star $^\star \mathcal G$.}  with $\mathcal C_\mu$ the {\it axial} vector potential of the 
 $U_d(1)$ group, and the $\dots$ denote the corresponding electric and magnetic current $J_{e,m}$ source terms, as in \eqref{lagZ}, which however we shall set to zero in our formalism, as we shall discuss below. The matter lagrangian, invariant under $U_{\rm em}(1)$, is given, for concreteness, by that of a complex (charged) {\it scalar} field of charge $q_e$, coupled to the electromagnetic potential $\mathcal A_\mu$ with a $U_{\rm em}(1)$ gauge-invariant potential $\mathcal V(\phi^\dagger\,\phi)$ describing scalar self interactions.
 \begin{align}\label{scalar}
 \mathcal L_{\rm matter} (\phi^\dagger, \phi, \mathcal A) = \Big[(\partial_\mu + i q_e \, \mathcal A_\mu )\phi\Big]^\dagger \, (\partial_\mu + i q_e \, \mathcal A_\mu )\phi  - \mathcal V(\phi^\dagger\, \phi)\,.
 \end{align}
The potential can be taken to be that of Higgs (Mexican hat) 
\begin{align}\label{higgs}
\mathcal V(\phi^\dagger\, \phi) = -\mu^2 \phi^\dagger \, \phi + \lambda (\phi^\dagger \, \phi)^2, \quad \, \mu \in  \mathbb R, \lambda > 0\,,
\end{align}
if we want spontaneous symmetry breaking of the electromagnetic $U_{\rm em}(1)$, but, in general, we do not restrict ourselves to that case. 

On representing the $\delta$-functional constraint in the path-integral as the following limit:
\begin{align}\label{deltarep}
\delta\Big[(F_{\alpha\beta} + \widetilde{\mathcal{G}}_{\alpha\beta})\,(F^{\alpha\beta} + \widetilde{\mathcal{G}}^{\alpha\beta})\Big] =  \lim_{\xi \to 0} \exp\Big(-i \frac{1}{\xi^2} \, \int d^4 x \, 
(F_{\alpha\beta} + \widetilde{\mathcal{G}}_{\alpha\beta})\,(F^{\alpha\beta} + \widetilde{\mathcal{G}}^{\alpha\beta}) \Big)
\end{align}
and going back to Minkowski space by analytic continuation, the effective Lagrangian stemming from the 
constrained path integral \eqref{cPI}
 in agreement with the solutions \eqref{gensol}, \eqref{gensol2}, 
reads~\cite{Zwanziger:1970hk}:
{\small \begin{align}\label{lag}
{\mathcal L}^{\rm eff} &= -\Big(\frac{1}{4q_e^2} + \frac{1}{q_e^2\xi^2}\Big)\, F_{\mu\nu}(A)F^{\mu\nu}(A) - \Big(\frac{1}{4q_m^2} + \frac{1}{q_m^2\xi^2}\Big)\,  \mathcal G_{\mu\nu} (C) \, \mathcal G^{\mu\nu}(C) - \frac{2}{q_e\, q_m\, \xi^2} \, F_{\mu\nu}(A) \, \widetilde{\mathcal G}^{\mu\nu}(C) + 
\mathcal L^\prime_{\rm matter}(\phi^\dagger, \phi, A) + \dots\,, \nonumber \\
&A_\mu \equiv q_e \, \mathcal A_\mu\,, \quad C_\mu \equiv q_m\, \mathcal C_\mu\, \quad \Rightarrow \, \quad
\mathcal L^\prime_{\rm matter}(\phi^\dagger, \, \phi, \, A) \equiv \mathcal L_{\rm matter}(\phi^\dagger, \, \phi, \, \mathcal A \to A, \, q_e \to 1)\,,
\end{align}}
with the $\dots $ denoting, as already mentioned, the source $J_{e,m}$ terms that we shall set formally to zero in our formalism, as explained below. The rescaling of the gauge potentials is dictated by the lattice model which we shall discuss in the next section \ref{sec:lattice}.
In our work we shall go beyond the classical effective field theory of \cite{Zwanziger:1970hk} by treating the parameter 
$\xi$ as a phenomenological coupling, away from its formal value $\xi \to 0$, stemming from the $\delta$-function representation \eqref{deltarep}. The reader should notice that, for any finite $\xi^2 < \infty$, the potentials $A$ and $C$ interact via the mixed Chern-Simons (CS) terms in the Lagrangian \eqref{lag}:
\begin{align}\label{CSterms}
{\mathcal L}^{\rm eff}_{\rm CS} = -\frac{2}{q_e\, q_m\, \xi^2} \, F_{\mu\nu}(A) \, \widetilde{\mathcal G}^{\mu\nu}(C)\,.
\end{align}
However, for regular $A$ and $C$, whose field strengths satisfy the ordinary Bianchi identities ({\it cf.} \eqref{nbianchi} below), a partial integration reveals that the terms \eqref{CSterms} are total derivatives, and thus do not contribute to the equations of motion for standard quantum field theories whose fields and their derivatives vanish at space-time infinity. As we shall argue, however, in this article, the presence of the monopole source induces singular potentials $A$ and $C$ in the quantum theory, and thus the mixed CS terms are non trivial. As we shall see in section \ref{sec:lattice}, these interactions play a crucial r\^ole in the phase diagram of the (strongly coupled) gauge theory \eqref{lag}, which we study here using Lattice techniques.

The reader should notice that the product of the coupling constants $q_e \, q_m$ does not necessarily obey the Dirac quantisation condition \eqref{dirac}. To this end, we need to determine the correct interaction of the matter with the monopole background, which sources a radial magnetic field:
\begin{align}\label{monmagn}
\vec{B}^{\rm sing}_{\rm mon} = q_m\, \frac{\hat r}{r^2}\,
\end{align}
where $\hat r$ is the unit vector along the direction of the radial spatial coordinate $r$, and   $q_m = \frac{\ell}{n\,\alpha} \, e$, $\ell  \in \mathbb Z$, is the magnetic charge obeying the Dirac quantisation condition \eqref{dirac} ({\it cf.} \eqref{magch}), where $q_e=n\, e$,  $n \in \mathbb Z$, is the electric charge of the field $\phi$ (for the standard charged Higgs $n=1$). The magnetic field is singular at the origin of the monopole $r=0$. We may represent this singularity as corresponding to a singular background field strength:
\begin{align}\label{dualback}
F_{ij} (\mathcal A_\mu^{\rm back}) = F_{ij}^{\rm back} (\mathcal A_\mu^{\rm back}) \, \equiv \,   \epsilon_{ijk} \, B^{\rm sing}_{\rm mon\, k}\,, \quad i,j,k = 1,2,3 \,\,({\rm spatial~indices})\, \, \Rightarrow \,\, F_{ij}^{\rm back} (A_\mu^{\rm back}) \, \equiv \,  \epsilon_{ijk} \, q_e\, B^{\rm sing}_{\rm mon\, k}\,, 
\end{align}
with $\epsilon_{ijk}$ the Euclidean 3-space Levi-Civita symbol. In \eqref{dualback}, we took into account that $\mathcal A^{\rm back}_\mu = \frac{1}{q_e}\, A_\mu^{\rm back}$.

The background \eqref{dualback} can be written in a covariant form, with (3+1)-dimensional indices, by assuming that 
all the rest of the indices of the background tensor $F_{\mu\nu}^{\rm back}$, involving $\mu=0$ and/or $\nu=0$ yield zero. We then treat  the field strength of $F_{\mu\nu} (A)$
as involving {\it both} singular and non-singular field configurations of the potential $A_\mu$, which is quantised in a path integral. As we shall show in section \ref{sec:lattice}, and mentioned above, the singular configurations of the gauge field $A$ are a consequence of the presence of the magnetic monopole source, and imply non-trivial mixed CS terms \eqref{CSterms}.

Let us see now what amendments we need to make to the Lagrangian \eqref{lag} so as to reproduce the classical equations \eqref{maxwell}. To this end, we first couple the model Lagrangian density \eqref{lag} to the background \eqref{monmagn} by setting $J_e^\mu=J_m^\mu=0$, $\mu=0,~\dots,~3$, and introducing the monopole background via the aforementioned background term $A_\mu^{\rm back}$ (see \eqref{dualback}) of the electromagnetic potential $A_\mu$, as follows:
\begin{align}\label{lag2}
&{\mathcal L}^{\rm eff} = - \frac{1}{4q_e^2} \Big(1 + \frac{4}{\xi^2}\Big)\, F_{\mu\nu}(A)F^{\mu\nu}(A) - 
\frac{1}{4q_m^2} \Big( 1 + \frac{4}{\xi^2}\Big)\,\mathcal G_{\mu\nu} (C) \, \mathcal G^{\mu\nu}(C) \nonumber \\
&- \frac{2}{q_e\, q_m\, \xi^2} \, (F_{\mu\nu}(A) + \chi {F}^{\mu\nu \,\rm back} (A^{\rm back}))\, \widetilde{\mathcal G}^{\mu\nu}(C) + 
\mathcal L^\prime_{\rm matter}(\phi^\dagger, \phi, A)\,,
\end{align}
where $F^{\mu\nu\, {\rm back}} =0 $ if any of $\mu, \nu=0$, otherwise it assumes the value \eqref{dualback}. At this point, 
let us make the important remark that by shifting the $F^{\mu\nu}$ by the background ${F}^{\mu\nu \rm back} (A^{\rm back})$ in the Maxwell-like kinetic terms of the Lagrangian \eqref{lag2}, does not affect the equations of motion for the configuration of the magnetic monopole background \eqref{monmagn}, as one can readily see (the so-resulting constant Lagrangian term ${F}^{\mu\nu \rm back} (A^{\rm back})\, {F}_{\mu\nu \, \rm back} (A^{\rm back})$ is irrelevant from the path integral point of view). 
This explains why in \eqref{lag2} we coupled only the dual tensor $\widetilde G_{\mu\nu} $ to the background \eqref{monmagn} via a CS coupling. This coupling term is parity invariant, because the potential $C_\mu$ is axial. We also notice that the scalar matter $\phi$ in \eqref{lag2} couples only to the potential $A_\mu$ and {\it not} to the monopole background.\footnote{For completeness, we mention that such a situation also characterises the Lattice Abelian gauge models of \cite{heller} in external fields.}

It is important to notice that the coefficient $\chi$ is to be determined by the requirement that in the limit $\xi \to 0$ one recovers, in the classical limit, the magnetic monopole model of \cite{Zwanziger:1970hk}, as we shall discuss below. In fact, if $\chi=1$ the theory, as we shall demonstrate below, is inconsistent.
To this end, we discuss now the classical Euler-Lagrange equations for the gauge fields $A_\mu$, $C_\mu$.

The equations of motion with respect to the field $A_\mu$, stemming from  \eqref{lag2} 
read (attention of the reader is drawn to the fact that the background (source) term $F_{\mu\nu}^{\rm back}(A^{\rm back})$ is {\it not} varied with respect to the (non singular) field $A_\mu$):
\begin{align}\label{lageqA}
&\frac{1}{q_e^2} \Big(1 + \frac{4}{\xi^2}\Big)\, \partial^\mu F_{\mu\nu}(A)  = - \frac{1}{q_e \, q_m}\, \frac{4}{\xi^2}  \,
\partial^\mu \widetilde{\mathcal G}_{\mu \nu}(C)  +  i \phi^\dagger \partial_\nu \phi - i (\partial_\nu \phi )^\dagger \, \phi - 2 A_\nu \phi^\dagger \, \phi  \quad \Rightarrow \nonumber \\
& \Big(1 + \frac{4}{\xi^2}\Big)\, \partial^\mu F_{\mu\nu}(\mathcal A)  = - \frac{4}{\xi^2}  \,
\partial^\mu \widetilde{\mathcal G}_{\mu \nu}(\mathcal C)  + q_e\Big( i \phi^\dagger \partial_\nu \phi - i (\partial_\nu \phi )^\dagger \, \phi - 2 q_e \mathcal A_\nu \phi^\dagger \, \phi \Big)\,,
\end{align}
where in the second line we took into account the rescaling \eqref{lag} of the gauge potentials. The last equation can be conveniently written as 
\begin{align}\label{lageqA2a}
 \partial^\mu F_{\mu\nu}(\mathcal A)  =   q_e\Big( i \phi^\dagger \partial_\nu \phi - i (\partial_\nu \phi )^\dagger \, \phi - 2 q_e \mathcal A_\nu \phi^\dagger \, \phi \Big) - \frac{4}{\xi^2}\, \Big(\partial^\mu F_{\mu\nu}(\mathcal A)  +  \,
\partial^\mu \widetilde{\mathcal G}_{\mu \nu}(\mathcal C) \Big) \,,
\end{align}
Thus, in the limit $\xi \to 0$, which corresponds to the classical theory of \cite{Zwanziger:1970hk}, Eq.~\eqref{lageqA2a}  leads to a finite standard result for the electric current of scalar electrodynamics (SQED) if an only if the constraint
\eqref{solconstr} is satisfied when $\xi \to 0$, which in this context implies that:
\begin{align}\label{constr2}
\lim_{\xi \to 0} \partial^\mu \Big(F_{\mu\nu}(\mathcal A)  + \widetilde{\mathcal G}_{\mu \nu}(\mathcal C) \Big) = 0 = \lim_{\xi \to 0}
\partial^\mu \Big(\widetilde{F}_{\mu\nu}(\mathcal A)  + {\mathcal G}_{\mu \nu}(\mathcal C) \Big) \,.
\end{align}
Hence, the $\xi \to 0$ limit of the $1/\xi$ terms in \eqref{lageqA2a} is identically zero, leaving only the $\xi$-independent terms, leading to \eqref{lageqA2}: 
\begin{align}\label{lageqA2}
\lim_{\xi \to 0} \partial^\mu F_{\mu\nu}(\mathcal A)  =  q_e \Big( i \phi^\dagger \partial_\nu \phi - i (\partial_\nu \phi )^\dagger \, \phi - 2 A_\nu \phi^\dagger \, \phi \Big) \equiv J^e_\nu \,,
\end{align}
where the right-hand side is the standard electric current of the scalar electrodynamics, corresponding to the Noether current $J^{\rm Noether}_\nu$  of the associated global version of the electromagnetic gauge symmetry $U_{\rm em}$. 
 For any finite $\xi \ne 0$, on  writing, 
\begin{align}\label{Aeq}
\partial^\mu F_{\mu\nu} (\mathcal A) = J^{e\,\rm eff}_\nu [\mathcal C, \phi,\, \partial \phi] \equiv  
- \frac{4}{4 + \xi^2}  \, \partial^\mu \widetilde{\mathcal G}_{\mu \nu}(\mathcal C)  + q_e \, \Big(1 + \frac{4}{\xi^2}\Big)^{-1}\, \Big( i \phi^\dagger \partial_\nu \phi - i (\partial_\nu \phi)^\dagger \, \phi - 2 q_e \mathcal A_\nu \phi^\dagger \, \phi \Big)\,,\,,
\end{align}
we observe that the effective ``current'' $J_\nu^{\rm eff}[\mathcal C, \,\phi, \,\partial \phi]$ receives also $\xi$-dependent contributions from the dual field strength of the $U_d(1)$ potential $\mathcal C$.

On the other hand, the equations of motion with respect to the field $C_\mu$, stemming from \eqref{lag2}, read:
\begin{align}\label{lageqB}
 \Big(1 + \frac{4}{\xi^2} \Big)\, \partial^\mu \mathcal{G}_{\mu\nu}(C) &= - \frac{2q_m}{q_e\, \xi^2}  \, \epsilon_{\mu\nu\rho\sigma} \, \partial^\mu \Big(F^{\rho\sigma} + \chi F^{\rho\sigma\, {\rm back}}(A^{\rm back})\Big)\, \quad \Rightarrow  \nonumber \\
\Big(1 + \frac{4}{\xi^2} \Big)\, \partial^\mu \mathcal{G}_{\mu\nu}({\mathcal C}) &= -  \frac{4}{\xi^2} \, \partial^\mu \Big(\widetilde{F}_{\mu\nu} ({\mathcal A})+ \chi \widetilde{F}_{\mu\nu\, {\rm back}}({\mathcal A}^{\rm back})\Big)\,, 
\end{align}
which can be conveniently written as:
\begin{align}\label{lageqBxi}
\partial^\mu \mathcal{G}_{\mu\nu}({\mathcal C}) = - \frac{4}{\xi^2} \, \partial^\mu \Big(\mathcal{G}_{\mu\nu}({\mathcal C}) 
+ \widetilde{F}_{\mu\nu} ({\mathcal A}) \Big) - \frac{4\chi}{\xi^2}\, \partial^\mu \widetilde{F}_{\mu\nu\, {\rm back}}({\mathcal A}^{\rm back})\,. 
\end{align}
The reader should have noticed that in arriving at both \eqref{lageqA} and \eqref{lageqB} 
we did {\it not} assume {\it a priori}  the Bianchi identities for the regular tensors $F_{\mu\nu}(A)$ and $\mathcal{G}_{\mu\nu}(C)$, that is, in our context we have:
\begin{align}\label{nbianchi}
\epsilon_{\mu\nu\rho\sigma}\, \partial^\mu\, F^{\rho\sigma}(A) \, =  2 \, \partial^\mu \widetilde{F}_{\mu\nu}(A) \ne 0 \,,  \quad 
\, \epsilon_{\mu\nu\rho\sigma}\, \partial^\mu\, \mathcal{G}^{\rho\sigma}(C) = 2 \, \partial^\mu \widetilde{\mathcal{G}}_{\mu\nu}(C) \ne 0\,.
\end{align}
 This will be understood from our subsequent Lattice analysis, which indicates that the coupling of the monopole background to the system of the gauge potentials $\mathcal A, \mathcal C$ induces monopole-like singularities 
in these fields, which invalidate the corresponding Bianchi identities.\footnote{It should also be noticed that, if the fields $\mathcal A, \mathcal C$  satisfied the Bianchi identities, the pertinent Euler-Lagrange equations would decouple, which is incorrect, as it does not correspond to the theory of \cite{Zwanziger:1970hk}.}

In similar spirit to the case of the $A$-field equations \eqref{lageqA2a}, in the limit $\xi \to 0$, the classical constraint \eqref{solconstr}, or equivalently \eqref{constr2} in this context, 
should be valid, which, implies that the $1/\xi^2$-dependent first term on the right-hand side of \eqref{lageqBxi} vanishes identically when $\xi \to 0$. Thus, on account of the fact that for the singular background \eqref{monmagn} we have:
\begin{align}\label{bianchback}
\epsilon_{\mu\nu\rho\sigma} \, \partial^\mu \, F^{\rho\sigma\, {\rm back}}(\mathcal A^{\rm back}) \ne 0\, ,
\end{align}
one obtains a finite non-trivial result for the $\mathcal C$-field equation when $\xi \to 0$, if an only if $\chi/\xi^2$ is $\xi$-independent. On account of \eqref{maxwell} and \eqref{gensol}, the right-hand side of the equations in \eqref{lageqB} in the limit $\xi \to 0$, would then define,  up to a crucial  minus sign, the magnetic current in the effective theory of \cite{Zwanziger:1970hk}:
\begin{align}\label{magcurr}
 J^{\rm mag}_\nu (\mathcal A^{\rm back}) \equiv  \,  \frac{2\chi}{\xi^2}\, \epsilon_{\mu\nu\rho\sigma} \, \partial^\mu F^{\rho\sigma\, {\rm back}}(\mathcal A^{\rm back})  = \partial^\mu \widetilde{F}_{\mu\nu}^{{\rm back}}(\mathcal A^{\rm back})  \,,
 \end{align}
upon the normalization
\begin{align}\label{xicoeff}
\chi =  \frac{\xi^2}{4}\,.
\end{align}
We stress once again that the definition \eqref{magcurr} , \eqref{xicoeff}, is consistent with \eqref{maxwell} and the constraint \eqref{solconstr} (or, equivalently, \eqref{constr2}). In this way, in the limit $\xi \to 0$, one recovers the standard monopole theory of \cite{Zwanziger:1970hk} with respect to the original potentials $\mathcal A_\mu$ and $\mathcal C_\mu$.

For generic $\xi \ne 0$, the effective magnetic current, which would depend on both, the fluctuations $\mathcal A$ of the background $\mathcal A^{\rm back}$ and $\mathcal A^{\rm back}$ itself, would read:
\begin{align}\label{magcurrxi}
 J^{\rm mag\, eff}_\nu (\mathcal A, \, \mathcal A^{\rm back}) =  \,  \frac{2}{4 + \xi^2} \, \Big(\epsilon_{\mu\nu\rho\sigma} \, \partial^\mu F^{\rho\sigma}(\mathcal A) + \chi \, \epsilon_{\mu\nu\rho\sigma}\, \partial^\mu F^{\rho\sigma\, {\rm back}}(\mathcal A^{\rm back})\Big)  \equiv  \frac{4}{4 + \xi^2} \Big(\partial^\mu {\widetilde F}_{\mu\nu} (\mathcal A) + \chi \, \partial^\mu {\widetilde F}^{\rm back}_{\mu\nu} (\mathcal A^{\rm back})\Big) \,,
 \end{align}
It is important for the reader to notice that, as a consequence of the singular nature \eqref{monmagn} of the magnetic field background at $r=0$ (see also \eqref{dualback}), the quantity $J_\nu^{\rm mag\, eff}$  is non trivial for any value of $|\xi| < +\infty$. 
 
Assuming static fields, as standard in the magnetic monopole case, Eq.~\eqref{lageqBxi} in the limit $\xi \to 0$, can be expressed as:
\begin{align}\label{magfield}
\lim_{\xi \to 0} \partial^i {\mathcal G}_{i0} (\mathcal C) = - \rho_m =  - \vec{\nabla} \cdot \vec{B}^{\rm sing}_{\rm back} \ne 0\,,
\end{align} 
with $\rho_m$ the magnetic charge density, which yields the standard Maxwell equation in the presence of magnetic monopoles, as it should be expected.

Because of this, the presence of magnetic monopoles would prevent superconducting properties of the Higgs vacuum. 
Indeed, to have superconductivity one requires a vanishing electric field, 
\begin{align}\label{elec}
\vec E=0\,,
\end{align}
and a non-zero magnetic field $\vec B \ne 0$, which is expelled from the bulk of the superconductor, up to a penetration depth (Meissner effect). In the standard SQED, this depth is inversely proportional to $|qv|$ (which is the same as the photon mass in the Higgs phase of SQED, that is, in the phase of spontaneously broken symmetry, where the scalar matter field acquires a non-trivial vacuum expectation value ($<\phi>=<\phi^\dagger>=\frac{v}{\sqrt{2}} \ne 0$)).  

The vanishing $(\vec E)^i=F^{0i}$ implies in our effective theory that, in the same limit $\xi \to 0$, the temporal component of eq. \eqref{lageqA2} yields in the Higgs phase:
\begin{align}\label{rho}
\rho = + q_e^2\, v^2 \, A_0
\end{align}
where $\rho$ is the electric charge density, and we have set $\partial^i F_{0i} = J_0 = \rho$. On account of \eqref{elec} we thus have $A_0=\rho=0$.

On the other hand, the spatial component of \eqref{lageqA2} (in the limit $\xi \to 0$) yields the current:
\begin{align}\label{current}
\lim_{\xi \to 0} \partial^j F_{ji}= J^e_i = + q_e^2 v^2 \, A_i .
\end{align} 
In the absence of a magnetic charge $\rho_m$, this would be the standard London equation leading to the Meissner effect with the aforementioned properties. But when $\rho_m \ne 0$ ({\it cf.} Eq.~\eqref{magfield}), the situation changes, since 
upon standard manipulations one would arrive at the following equation for the magnetic field intensity
$\vec{\nabla} \cdot (\vec{\nabla} \cdot \vec B) - \nabla^2 \vec B + q_e^2 v^2 \vec B =0$. The non-zero term $\vec{\nabla} \cdot (\vec{\nabla} \cdot \vec B) \ne 0$ spoils the Meissner effect in the case of magnetic monopole backgrounds.

We also remark, for completion at this stage, that in the case $\xi \ne 0$, the effective ``electric'' Noether current of SQED \eqref{Aeq}
receives contributions from the dual field $C$, which also spoils any superconducting properties of the Higgs-phase ground state:
\begin{align}\label{currentxi}
\partial^j F_{ji}\Big|_{\phi=<\phi>=v/\sqrt{2}}= J^{\rm Noether, ``electric''}_i \Big|_{\phi=<\phi>=v/\sqrt{2}}= -\frac{4}{4 + \xi^2} \partial^j \widetilde{\mathcal G}_{ji} + \frac{q_e^2 v^2}{1 + \frac{4}{\xi^2}}A_i  \,, \quad \xi \ne 0\,.
\end{align} 
From this expression we observe that only in the absence of monopoles, i.e. in the case $\xi \to \infty$, 
(for which the field $C_\mu$ decouples in the path integral), eq.~\eqref{currentxi} yields the 
London equation of superconductivity (and the associated Meissner effect), which corresponds to the broken $U_{\rm em}(1)$ phase of the scalar electrodynamics, according to standard arguments. 

In general, one should put  the Lagrangian \eqref{lag2} on the Lattice, keeping the quantities $q_e, \xi^2$ and $q_m$ {\it arbitrary}. According to the above analysis, the classical monopole solution of 
\cite{Zwanziger:1970hk} should correspond to a fixed point $\xi \to 0$ at which the constraint \eqref{solconstr} is realised. This is the main focus of the current article, which is discussed from the next section onwards. 

Before proceeding to the lattice non-perturbative study, we should  stress once again that for any $\xi < \infty$, the effective Lagrangian \eqref{lag2} contains CS terms that mix the two gauge potentials $\mathcal A$ and $\mathcal C$. We remind the reader that such terms arise initially (for $\xi \to 0$) from implementing the constraint \eqref{solconstr} of \cite{Zwanziger:1970hk} in the path integral. At this point we should mention for completeness that
such terms arise naturally (upon appropriate constraints in the gauge couplings) in gauged supergravity model with $U(1) \times U^\prime (1)$ gauge groups~\cite{sugra}.\footnote{We thank F. Farakos for a useful discussion on this point.} Also, the analogue of such mixed terms arise in (2+1)-dimensional $U(1)_{\rm em} \times U(1)_{\rm stat}$ gauge models of  high-temperature superconductors~\cite{dorey}, in which the $U(1)_{\rm em}$ is the group of the electromagnetism, while 
the group $U(1)_{\rm stat}$ represents the ``statistical'' gauge group, associated with the attractive force generated by the anyon statistics of the excitations of these planar models.

Although below we shall mainly concentrate on the lattice (non-perturbative) version of
the Lagrangian \eqref{lag2}, nonetheless,  
in order to demonstrate the special r\^ole of the coupling between the electromagnetic and dual $U(1)$ via the CS-type kinetic mixing  \eqref{CSterms}, $F_{\mu\nu}(A) \, \widetilde{\mathcal G}^{\mu\nu}(C)$, we shall also consider in section  \ref{sec:FAFC}, for comparison, an alternative Lattice action with an ordinary kinetic mixing between the two $U(1)$'s in the absence of the interaction \eqref{CSterms}:
\begin{align}\label{ordinary}
\mathcal L^{\rm eff} \ni \frac{2\, \lambda}{q_eq_m} \, F_{\mu\nu} \, \mathcal G^{\mu\nu}\,,
\end{align}
where $\lambda \in \mathbb  R$ a real coupling parameter (in our case this is fixed as in \eqref{FAFC}, section \ref{sec:FAFC}). For our purposes here we do not ascribe any potential physical significance to such a term, as we  simply consider it to differentiate its effects from those of the 
dual mixing term in \eqref{lag2}, as far as the symmetry breaking  patterns and the r\^ole of the monopole configuration are concerned, as we shall discuss below. We do note, though, for completeness, that kinetic mixings between U(1) of the form \eqref{ordinary}
are considered in massive-photon models of (millicharged) Dark matter~\cite{mcdm}, upon inclusion of matter fermions coupled to both $U(1)$'s, in the phase where the non electromagnetic $U(1)$ is spontaneously broken. However, in our case, as already mentioned below 
 \eqref{relAB}, due to the constraint \eqref{relAB} (or \eqref{solconstr}), the vector potential $\mathcal C_\mu$ of the dual $U_d(1)$ is an {\it  axial (pseudo)vector}, which is not the case in \cite{mcdm}. In the current article we shall not consider such implications, postponing the pertinent discussion for the future.

\section{The Lattice Model}\label{sec:lattice}

The action representing the expression (\ref{lag2}) on the lattice reads: 
\begin{align} S &= \f{\beta_A}{2} \left(1 +\f{4}{\xi^2}\right) \sum_x \sum_{1\le\mu<\nu\le 4} [F_{\mu\nu}^{A,latt}(x)]^2 + \f{\beta_C}{2} \left(1 +\f{4}{\xi^2}\right) \sum_x \sum_{1\le\mu<\nu\le 4} [F_{\mu\nu}^{C,latt}(x)]^2\nonumber \\&+\f{2\, \sqrt{\beta_A\, \beta_C}}{\xi^2} \sum_x \sum_{1\le\mu<\nu\le 4} \sum_{\rho,\sigma} \epsilon_{\mu\nu\rho\sigma} [(F^{A,latt}_{\mu\nu}(x)+ \chi \, F^{B,latt}_{\mu\nu}(x)) F^{C,latt}_{\rho\sigma}(x)] \nonumber \\ &+ \sum_x \Phi^*(x)\Phi(x) - \beta_h \sum_x \left[\sum_{1\le \mu\le 4} \Phi^*(x) U^A_{x \hat{\mu}} \Phi(x+\hat{\mu})\right] \nonumber \\ & + \beta_R \sum_x [\Phi^*(x)\Phi(x)-1]^2\,, \quad \chi=\frac{\xi^2}{4}\,.\label{act1}\end{align} 

The quantity $x$ is a collective index for all four coordinates. The lattice version of the field strengths is defined, for example, through: \be F_{\mu\nu}^{A,latt}(x)\equiv \theta^A_\mu(x) + \theta^A_\nu(x+\hat{\mu}) - \theta^A_\mu(x+\hat{\nu}) - \theta^A_\nu(x),\label{FA}\ee \be F_{\mu\nu}^{C,latt}(x)\equiv \theta^C_\mu(x) + \theta^C_\nu(x+\hat{\mu}) - \theta^C_\mu(x+\hat{\nu}) - \theta^C_\nu(x),\label{FC}\ee 
where $\theta^A_\mu(x) \equiv q_e a {\cal A}_\mu(x) = a\, A_\mu(x),\ \theta^C_\mu(x) \equiv q_m a {\cal C}_\mu(x) = a\,C_\mu(x)$ are the lattice versions of the potentials ${\cal A}_\mu(x),\ {\cal C}_\mu(x)$ at the position $x$ along the direction $\mu.$ 

The quantity $a$ is the lattice spacing and $q_e,\ q_m$ are the coupling constants. In the action (\ref{act1}) the symbol $x+\hat{\mu}$ denotes the first neighbour of site $x$ in the direction $\hat{\mu}.$ We also use the notation $U^A_{x \mu}=e^{i \theta^A_{x \mu}}$ for the link gauge field, which enters the interaction with the scalar field. The expression $F^{B,latt}_{\mu\nu}$ represents the monopole background field and the couplings $\beta_A = \f{1}{q_{e}^2},\ \beta_C = \f{1}{q_{m}^2}$ equal the inverse square couplings appearing in the continuum Lagrangian \eqref{lag2}. The $A$ field contains contributions from the monopole background (second line), as well as interaction with the scalar field (third line). Notice that the scalar field interacts only with the $A$ gauge field. Finally the constant $\beta_R$ corresponds to the quartic coupling of the scalar field. 

Writing $\Phi(x) = \rho(x) e^{i\phi(x)}$ and $U^A_{x \mu}=e^{i \theta^A_{x \mu}}$ the action becomes:
\begin{align} S &= \f{\beta_A}{2} \left(1 +\f{4}{\xi^2}\right) \sum_x \sum_{1\le\mu<\nu\le 4} [F_{\mu\nu}^{A,latt}(x)]^2 + \f{\beta_C}{2} \left(1 +\f{4}{\xi^2}\right) \sum_x \sum_{1\le\mu<\nu\le 4} [F_{\mu\nu}^{C,latt}(x)]^2\nonumber \\&+\f{2\sqrt{\beta_A\, \beta_C}}{\xi^2} \sum_x \sum_{1\le\mu<\nu\le 4} \sum_{\rho,\sigma} \epsilon_{\mu\nu\rho\sigma} [(F^{A,latt}_{\mu\nu}+ \chi F^{B,latt}_{\mu\nu}) F^{C,latt}_{\rho\sigma}] \nonumber \\ 
&+ \sum_x \rho^2(x) - \beta_h \sum_x \sum_{1\le \mu\le 4} \rho(x)\rho(x+\hat{\mu}) \cos[\phi(x+\hat{\mu})  + \theta^A_{x \mu} -\phi(x)] +\beta_R \sum_x [\rho^2(x)-1]^2.\label{actn}\end{align}
This is the precise form to be used in the lattice simulations. We have used the non-compact version of the lattice $U(1)$ models. This has been done to avoid monopoles arising from the compact formulation, such as the ones studied long ago ~\cite{degrand:1980}. 

\section{The Higgs model within and outside a monopole background}\label{sec:higgs}

A first step we take is to consider the model without the monopole background \eqref{monmagn}, i.e. with $\vec B^{\rm sing}_{\rm mon}=0.$ The point is that one should know what happens in this case, before investigating the influence of the monopole source. Of particular interest is the influence of the parameter $\xi.$ 

To proceed with the calculations we need the definitions of two observables, which will be used in the simulations: the angular part of the link variable 
\be V^{(0)} = \f{1}{N^4}\sum_{n_x,n_y,n_z,n_t=0}^{N-1} \cos[\phi(n_x+\hat{1},n_y,n_z,n_t) + \theta^A_{n_x,n_y,n_z,n_t,\hat{1}} - \phi(n_x,n_y,n_z,n_t)],\label{V1}\ee 
as well as the mean square \be R_2^{(0)} = \f{1}{N^4}\sum_{n_x,n_y,n_z,n_t=0}^{N-1} \Phi^\dagger(n_x,n_y,n_z,n_t) \Phi(n_x,n_y,n_z,n_t) = \f{1}{N^4}\sum_{n_x,n_y,n_z,n_t=0}^{N-1} \rho^2(n_x,n_y,n_z,n_t).\label{R21}\ee In the previous expressions we have denoted the lattice sites by the dimensionless integers $n_x,n_y,n_z,n_t.$ We have also used the notation $\hat{1}$ for direction $x,$ while $n_x+\hat{1}$ means the site next to $n_x$ in the direction $x.$ In $V^{(0)}$ we have used links in the $x$ direction; it would make no difference if we used any other spatial direction.

The quantity $R_2^{(0)}$ is the lattice version of the expectation value of $\phi^\dagger(x) \phi(x),$ that is a gauge invariant version of the vacuum expectation value (squared) of the scalar field. A large value of $R_2^{(0)}$ indicates the symmetry broken phase of the gauge-higgs theory. The symmetric phase is characterized by a small value of this indicator, typically of the order of one in lattice units.

The quantity $V^{(0)}$ is a gauge invariant lattice version of the expectation value of $\Big[(\partial_{\hat{1}} + i q_e \, \mathcal A_{\hat{1}} )\phi\Big]^\dagger \, \Big[(\partial_{\hat{1}} + i q_e \, \mathcal A_{\hat{1}})\phi\Big].$ A value of $V^{(0)}$ around one indicates the symmetry broken phase, while a small value for this quantity (around zero) corresponds to the symmetric phase of the model.

Then we calculate the statistical averages $<V^{(0)}>$ and $<R_2^{(0)}>$ versus $\beta_h$ for the values of the parameters set to $\beta_A=4.0,$ $\beta_C=0.25,\ \beta_R=0.001.$ We use Monte Carlo simulations to calculate these averages; typically we have performed 3000 lattice sweeps to thermalise the system and 2000 iterations to compute the averages (using one out of every five configurations). The details concerning the simulation of the mixed term are explained in the appendix.

\begin{figure}[ht]
\begin{center}
\includegraphics[scale=0.6,angle=0]{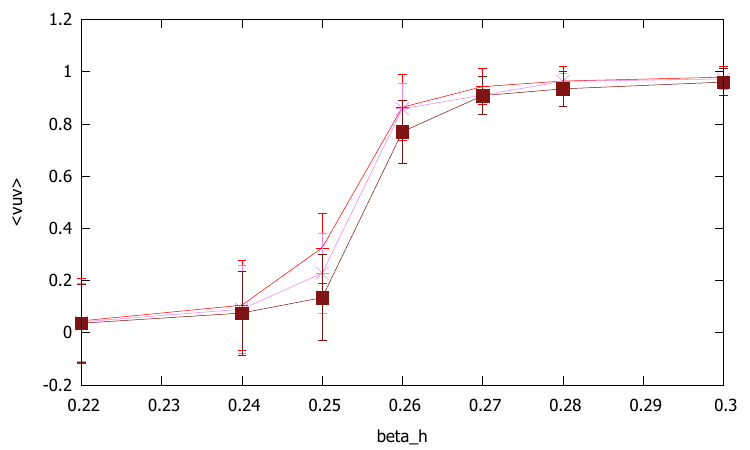}
\includegraphics[scale=0.6,angle=0]{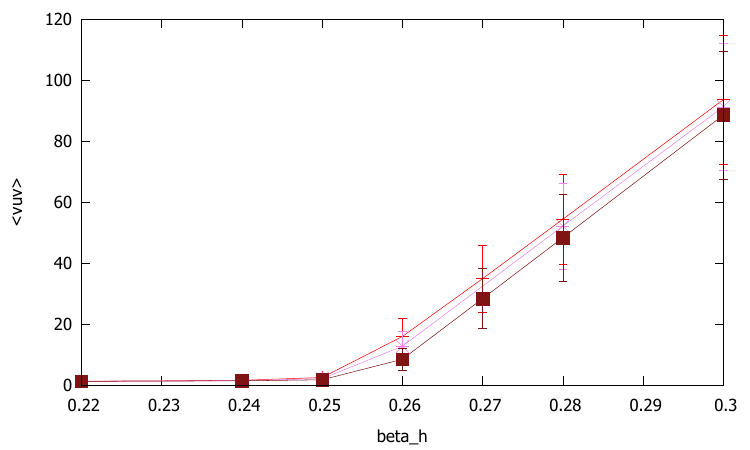}
\end{center}
\caption {Angular part $<V^{(0)}>$ of links (left) and mean measure squared $<R_2^{(0)}>$ of the scalar field (right) for $\xi=0.1,\ \xi=1.0$ and $\xi=10.0$ in the absence of the monopole background. The uppermost curves correspond to $\xi=0.1,$ while the lowest curves to $\xi=10.0.$ The relevant parameters read: $\beta_A=4.0,$ $\beta_C=0.25.$ $\beta_R=0.001.$} \label{link}
\end{figure}

Notice that $\beta_A \beta_C=1,$ which simplifies the lattice action. The results are shown in figure \ref{link}. We show the results for three quite different values of $\xi,$ namely $\xi=0.1,\ \xi=1.0$ and $\xi=10.0.$  The differences in $\xi,$ although quite large, do not have serious consequences and we see the expected transition from the symmetric to the broken phase. The phase transition lies between $\beta_h=0.24$ and $\beta_h=0.26$ for all values of $\xi$ considered. It is evident that the value $\beta_h=0.22$ corresponds to the symmetric phase, while $\beta_h=0.28$ corresponds to the broken phase. We will use these values for $\beta_h$ repeatedly in the sequel.

\subsection{Lattice Monopole background}

\begin{figure}[ht]
\begin{center}
\includegraphics[scale=0.6,angle=0]{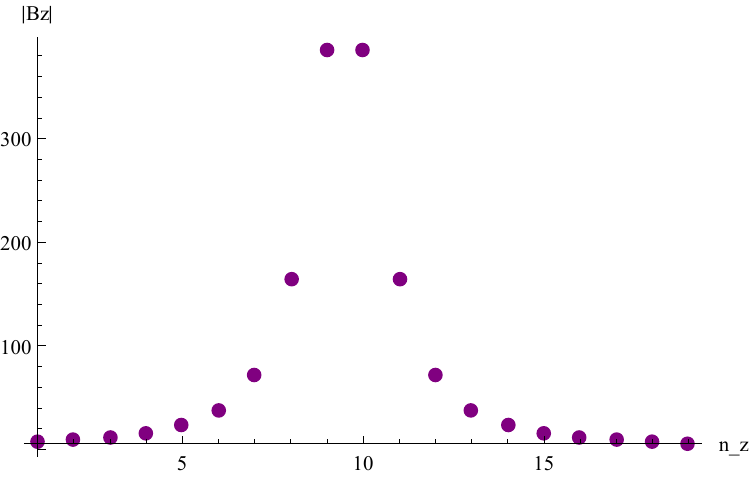}
\end{center}
\caption {Monopole background $B_z$ field.} \label{bkgr}
\end{figure}

Let us define the lattice monopole background. This will be derived from a source term corresponding to: 
\be\label{maglatt}
\vec{B} = B \f{\vec{n}-\vec{x}_M}{|\vec{n}-\vec{x}_M|^3},\ee 
where $\vec{n} \to (n_{x},n_{y},n_{z}),$ $\vec{x}_M \to (x_M,y_M,z_M)$ (the position of the monopole source) and $B$ is a dimensionless number. Throughout this work we will set $B=500$ and use a $N^4 = 20^4$ lattice. We suppose that the monopole position is at the center of the lattice: if the $n_x$ coordinate takes on the values $0,\ 1,\ \dots,\ N-1,$ the position of the monopole lies at $x_M=y_M=z_M=\f{N-1}{2}.$ This position, which does not correspond to any lattice site, has been chosen on purpose, to avoid infinities. We set the monopole field strength equal to zero if $|\vec{n}-\vec{x}_M|>D_{max},$ where $D_{max}=N-\f{1}{2},$ that is we suppose that there exists a sphere of anti-monopoles, which exactly cancels the monopole field at large distances. A copy of the monopole is constructed for each value of the remaining coordinate t. In figure \ref{bkgr} we depict the absolute value $|B_{z}|$ of the monopole background field strength along $z$ (or whichever other spatial direction) versus $n_z.$ We have set $n_x=n_{x0}\equiv\f{N}{2},\ n_y=n_{y0}\equiv \f{N}{2},$ so that: 
\be |B_z| = 500 \f{\left|n_z-\f{19}{2}\right|}{\left[\f{1}{2}+\left(z-\f{19}{2}\right)^2\right]^{3/2}}.\ee

An important consequence of the monopole background field is that the system is no more homogeneous, since various quantities depend on the distance from the position of the monopole.

To describe the observables to be measured in the sequel we define a line along $z,$ with fixed values of $n_x=n_{x0}\equiv\f{N}{2},\ n_y=n_{y0}\equiv \f{N}{2},$ while $n_z$ and $n_t$ run from $0$ to $N-1.$ This line will lie at the basis of our calculations. The line passes very near the monopole background core, when $n_z=\f{N}{2},$ while its extremities lie away from it. It has been constructed to illustrate the different behaviours at small and large distances from the core. The various regions are labelled by the $n_z$ dependence of the results. 

We study two observables, which have to do with the higgs sector, namely the angular part of the space-like links: 
\be V(n_z) = \f{1}{N}\sum_{n_t} \cos[\phi(n_{x0}+1,n_{y0},n_z,n_t) + \theta^A_{n_{x0},n_{y0},n_z,n_t,\hat{1}} - \phi(n_{x0},n_{y0},n_z,n_t)],\label{Vz}\ee 
as well as the mean square \be R_2(n_z) = \f{1}{N}\sum_{n_t} \Phi^\dagger(n_{x0},n_{y0},n_z,n_t) \Phi(n_{x0},n_{y0},n_z,n_t) = \f{1}{N}\sum_{n_t} \rho^2(n_{x0},n_{y0},n_z,n_t).\label{R2z}\ee  We have slightly changed our notation, in the sense that we write down explicitly the four coordinates, rather than using a collective index. The quantities $n_z,\ n_t$ take the values $0,\ 1,\ \dots,\ N-1.$ The coordinates $n_x$ and $n_y$ have been set to the values $n_{x0}=\f{N}{2}$ and $n_{y0}=\f{N}{2}.$ Both are local quantities. The variable $V(n_z)$ represents links between the sites $(n_{x0},n_{y0},n_z,n_t)$ and $(n_{x0}+1,n_{y0},n_z,n_t),$ where one may spot the scalar fields, while the gauge field between them is $U^A_{n_{x0},n_{y0},n_z,n_t,\hat{1}} = \exp\left[i \theta^A_{n_{x0},n_{y0},n_z,n_t,\hat{1}}\right].$ In these calculations we have used links in direction $\hat{1}.$ It would make no difference to use whichever other spatial direction; one might also form a sum over all three directions, but this would be more noisy. The variable $R_2(n_z)$ is simpler, since it does not involve any direction. 

\subsection{Symmetric phase}\label{sec:symm}

In the following we will concentrate on the value $\beta_h=0.22,$ which corresponds to the symmetric phase. The remaining parameters are set to $\beta_A=4.0,$ $\beta_C=0.25,$ $\beta_R=0.001$ and $B=500.$ We use $B=500$ in the most part of this paper.

\begin{figure}[ht]
\begin{center}
\includegraphics[scale=0.7,angle=0]{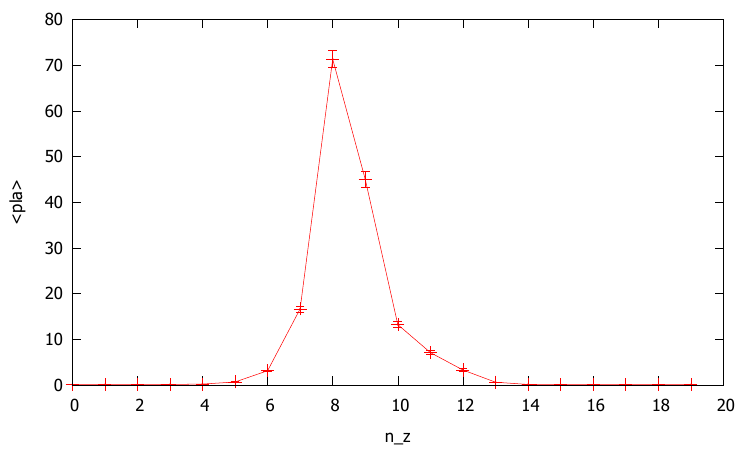}
\includegraphics[scale=0.7,angle=0]{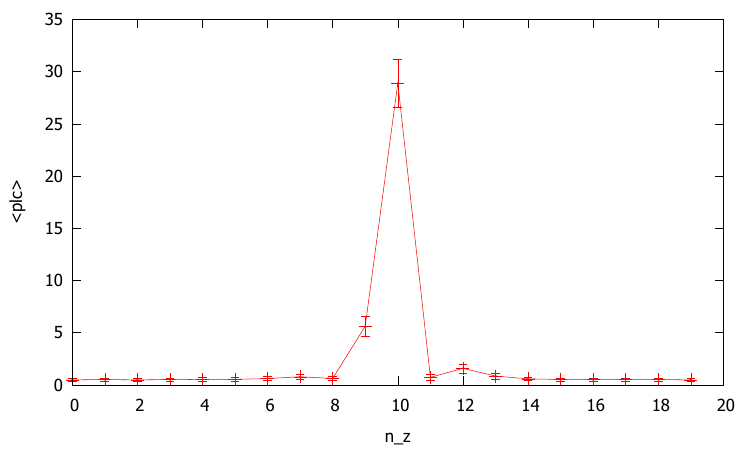}
\end{center}
\caption {Symmetric phase, $\xi=1.00.$ Plaquettes A (left panel) and C(right panel) in the symmetric phase versus $n_z.$ The parameters read: $\beta_A=4.0,$ $\beta_C=0.25,$ $\beta_h=0.22,$ $B=500$ and $\beta_R=0.001.$} 
\label{xi_bh1_symm}
\end{figure}

We start with the results for the A and C plaquettes (which are the lattice versions of the energy densities) and are defined through the expressions: 
\begin{align} 
Pl_A(n_z)=  \f{1}{N} \sum_{n_t} <[F_{12}^{A,latt}(n_{x0},n_{y0},n_z,n_t)]^2>,\ Pl_C(n_z)=  \f{1}{N} \sum_{n_t} < [F_{12}^{C,latt}(n_{x0},n_{y0},n_z,n_t)]^2>\,.
\end{align}
The field strengths $F_{12}^{A,latt}$ and $F_{12}^{C,latt}$ are defined in equations (\ref{FA}) and (\ref{FC}) respectively and $n_{x0},n_{y0}$ have been already defined. The results for $\xi=1.0$ are depicted in figures \ref{xi_bh1_symm}. We find that there is some structure around the position of the monopole background (around $n_z=9$) for both plaquettes. Simulations with $\beta_h=0,$ where the scalar field is decoupled from the gauge fields yield very similar results, implying that the influence of the monopole source on the gauge fields does not depend very much on the scalar field.

\begin{figure}[ht]
\begin{center}
\includegraphics[scale=0.7,angle=0]{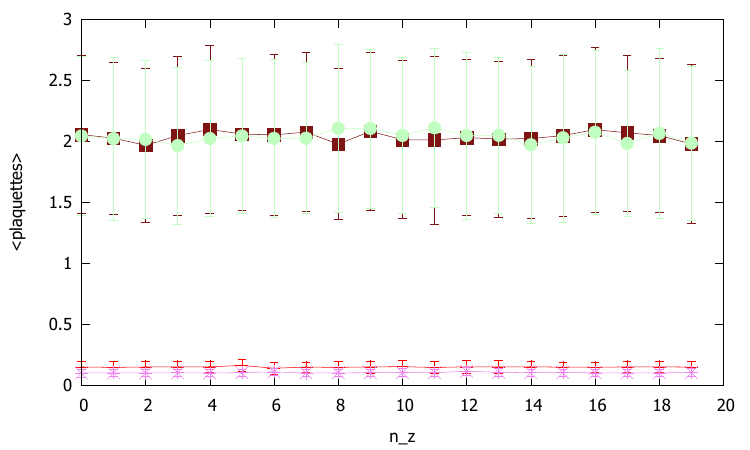}
\end{center}
\caption {Plaquettes A in the symmetric phase (intermediate line) and broken phase(lowest line) at $\xi=10.0$ Plaquettes C do not change as one moves from the symmetric to the broken phase and are represented by the uppermost lines. The parameters read: $\beta_A=4.0,$ $\beta_C=0.25,$ $B=500$ and $\beta_R=0.001;$ the symmetric phase is represented by setting $\beta_h=0.22,$ while the broken phase is reached by setting $\beta_h=0.28.$} 
\label{xi=10_pl}
\end{figure}

We have also performed the corresponding calculations for $\xi=10.0$ and we have found no similar peaks. We have also simulated models of non-compact $U(1)$ gauge fields with no coupling to scalar fields and no mixing terms and the results on the plaquettes at the values $\beta_A=4.0$ and $\beta_C =0.25$ are the same as the ones found in the previous model. In figure \ref{xi=10_pl} one may see plaquettes A and C at $\xi=10.0,$ in the symmetric (and the broken) phase. They do not exhibit any structure along the z direction and differ little from one another.

\begin{figure}[ht]
\begin{center}
\includegraphics[scale=0.6,angle=0]{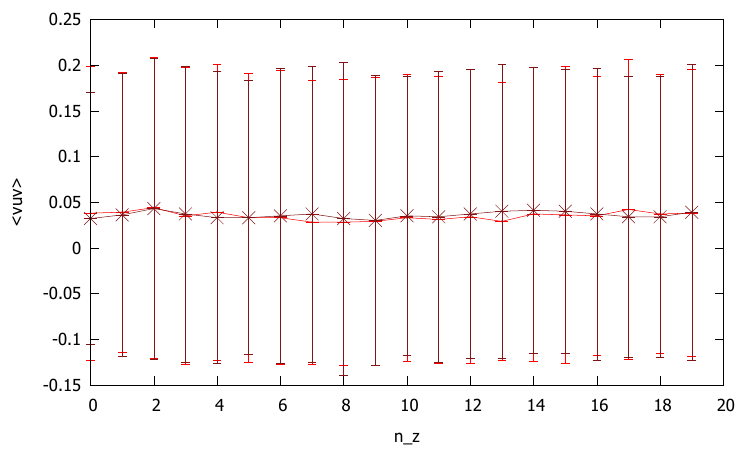}
\includegraphics[scale=0.6,angle=0]{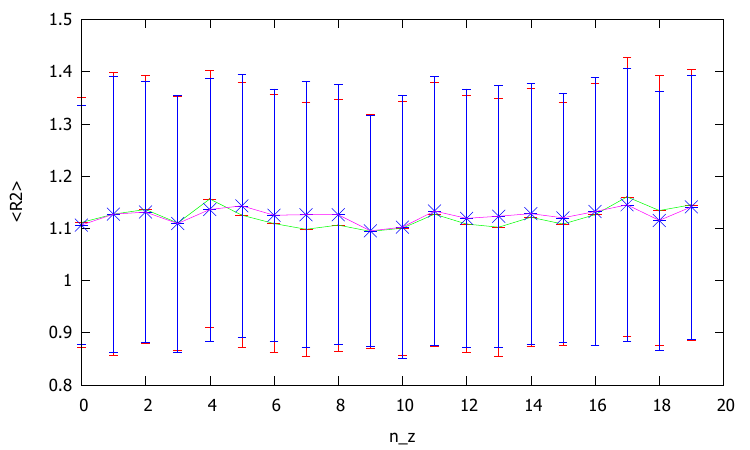}
\end{center}
\caption {Symmetric phase. Angular parts $<V(n_z)>$ of links in space-like directions (left) and $<R_2(n_z)>$ (right) versus $n_z.$ Couplings: $\beta_A=4.0,$ $\beta_C=0.25,$ $\beta_h=0.22,$ $B=500$ and $\beta_R=0.001.$ We depict results for $\xi=1.00$ and $\xi=10.00.$ } \label{linksymm}
\end{figure}

In figure \ref{linksymm} we depict the angular quantity $<V(n_z)>$ of links (left) and mean measure squared $<R_2(n_z)>$ (right) of the scalar field for $\xi=1.00$ and $\xi=10.00.$ Brackets denote statistical averages, as usual. No essential difference is detected between the two values of $\xi.$ Since there is a magnetic monopole background, one would expect some non-trivial dependence of various observables, such as $<V(n_z)>$ and $<R_2(n_z)>$ on $n_z.$ However, the $n_z$ dependence is barely seen, which means that, in the symmetric phase, no significant influence of the background on the system is detected. Thus, although the plaquettes take large values at the origin of the monopole background source, the link quantities do not detect any sign of the background. We have checked numerically that large values of $\xi$ render the system almost indistinguishable from a system with no coupling between the two gauge fields and the background magnetic field. 

\subsection{Broken phase}\label{sec:broken}

Now we examine the system in the broken phase, setting $\beta_h=0.28.$ The remaining variables read: $\beta_A=4.0,$ $\beta_C=0.25,$ $\beta_R=0.001$ and $B=500.$ In figure \ref{xi=10_pl}, apart from the results in the symmetric phase, one may also find the plaquettes A and C at $\xi=10.0$ in the broken phase. No structure appears in the z direction for these plaquettes.

On the other hand,we found that the corresponding results for the A and C plaquettes in the broken phase at $\xi=1.0$ are quite similar to the ones found for the symmetric phase. The difference between the symmetric and the broken phases show up in the behaviour of the quantities related to the links, rather than the plaquettes.

\begin{figure}[ht]
\begin{center}
\includegraphics[scale=0.6,angle=0]{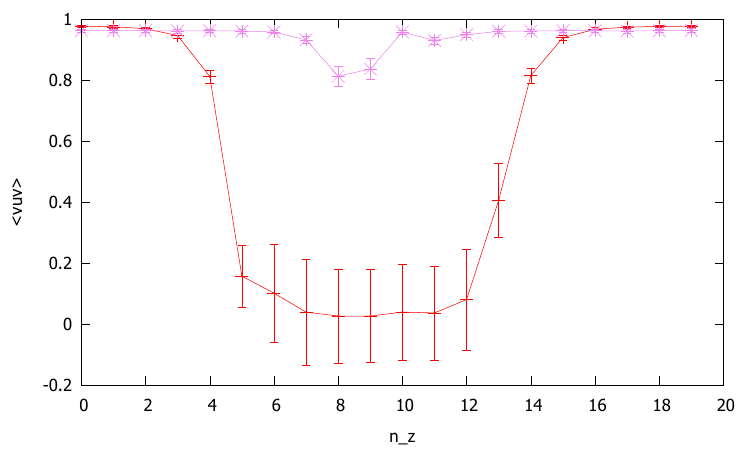}
\includegraphics[scale=0.6,angle=0]{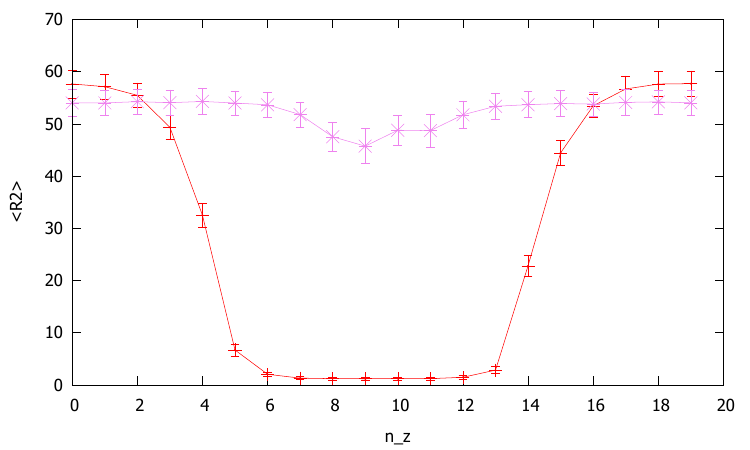}
\end{center}
\caption {Higgs phase. Angular parts $<V(n_z)>$ of links in space-like directions (left) and $<R_2(n_z)>$ (right) versus $n_z.$ Couplings: $\beta_A=4.0,$ $\beta_C=0.25,$ $\beta_h=0.28$ and $\beta_R=0.001.$ We depict results for $\xi=1.00$ and $\xi=10.00.$ The deepest wells correspond to $\xi=1.00.$ } \label{linkhiggs}
\end{figure}

We examine two values of $\xi,$ namely $1.00$ and $10.00,$ and depict the angular part $<V(n_z)>$ of links (left) and mean measure squared $<R_2(n_z)>$ (right) of the scalar field, as functions of $n_z$, in figure \ref{linkhiggs}. One may see results strikingly different from the ones corresponding to the symmetric phase. As already stated, for fairly large values of $\xi$ $(\xi=10.00$ for example), it is expected that the terms proportional to $1/\xi^2$ in the action are negligible, so that the model reduces to a simpler one, not involving the coupling of the two $U(1)'$s and the monopole background contribution. Thus no noticeable $n_z$ dependence of either $<V(n_z)>$ or $<R_2(n_z)>$ is expected for large $\xi.$ We have checked numerically that $\xi=10.00$ is a value, above which $\xi$ may be considered as large in the previous sense. A result supporting this is included in figure \ref{linkhiggs}, where it may be seen that, for $\xi=10.00,$ the quantities $<V(n_z)>$ and $<R_2(n_z)>,$ defined via equations (\ref{Vz}) and (\ref{R2z}), have a very mild $n_z$ dependence; this dependence becomes almost invisible for even larger values of $\xi.$ However, at $\xi=1.0,$ the $\xi-$dependent terms in the action come into play and the $n_z$ dependence is manifest, in the sense that a well develops around the core of the monopole source. In particular the measure of the scalar field approaches the value corresponding to the symmetric phase. This result is consistent with our understanding that in the monopole core the scalar field should take on very small values. Our results indicate that a monopole configuration has been created in the broken phase, through the couplings between the two $U(1)'$s and the background magnetic monopole (figure \ref{bkgr}), which is just an alternative form of the external magnetic current. The building up of the monopole configuration has to do with the mixed CS coupling in (\ref{actn}). The lattice appears to have a region where the system lies in the symmetric phase. Such a region corresponds to a three-dimensional sphere centered at the monopole origin.  The rest of the lattice remains in the broken phase, as shown in figure \ref{linkhiggs}. We will examine the $\xi$ dependence of this phenomenon in the next subsection.

\begin{figure}[ht]
\begin{center}
\includegraphics[scale=0.7,angle=0]{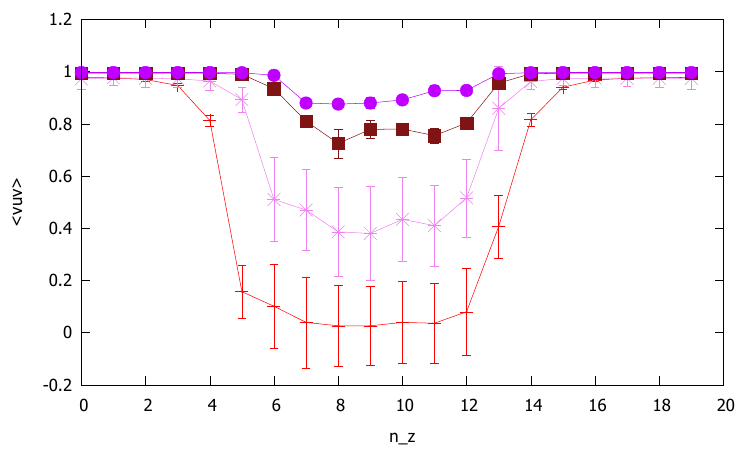}
\end{center}
\caption {Angular parts of links in space-like directions in the broken phase versus $n_z.$ The parameters read: $\beta_A=4.0,$ $\beta_C=0.25,$ $B=500$ and $\beta_R=0.001.$ The parameter $\xi$ is $1.00$ throughout, while $\beta_h$ takes the values $0.28$ (this corresponds to the deepest wells), $\ 0.30,\ 0.40$ and $0.50$ (the latter has the weakest $n_z$ dependence).} 
\label{plac}
\end{figure}

The monopole is present when $\beta_h$ is large enough to drive the system into the broken phase, however it is interesting to examine the role of this parameter in more detail. A particular scenario is that, if $\beta_h$ grows too large, it could be difficult to create a monopole out of the too massive scalar and gauge  fields, so the effects of the mixed CS-like terms may not be visible. We have chosen to use $\xi=1.00$ and examine the effect of $\beta_h$ on the angular parts of the links. We expect that, for this value of $\xi,$ the effects of $\beta_h$ will be easily visible. The angular parts of the links have been chosen, since they are good indicators for the existence of monopoles; in addition their absolute values do not exceed one, so it is easy to compare them for different values of $\beta_h,$ while the measure of the scalar field will vary widely and the comparison is not straightforward. The results may be seen in figure \ref{plac}. At $\beta_h=0.28,$ i.e. close to the phase transition towards the broken phase, we testify the appearance of deep wells, that is regions of symmetric phase in the middle of the  lattice, which remains at the broken phase away from the center. As $\beta_h$ increases to $0.30,$ the well is slightly less deep, while it becomes even more shallow when $\beta_h=0.40$ and $\beta_h=0.50.$ In the last case it appears that the whole lattice lies in the broken phase. In short, an increase of $\beta_h$ tends to undo the effect of the background source and its coupling to the $A$ fields. 

\subsection{Dependence on $\xi.$}\label{sec:xidep}

\begin{figure}[ht]
\begin{center}
\includegraphics[scale=0.9,angle=0]{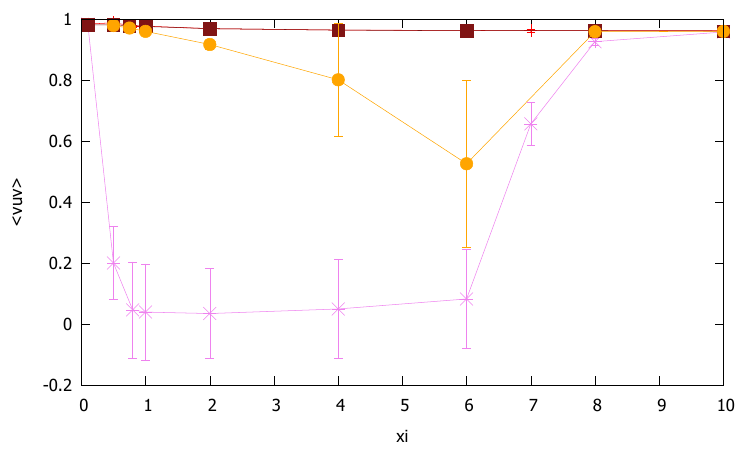}
\end{center}
\caption {Broken phase. Angular parts of links in space-like directions away from the core $<V(n_z=0)>$ (uppermost curve), $<V(n_z=10)>$ near the core for $B=500,$(lowest curve) and $<V(n_z=10)>$ near the core for $B=100,$(intermediate curve) in the broken phase versus $\xi.$ The parameters read: $\beta_A=4.0,$ $\beta_C=0.25,$ $\beta_h=0.28$ and $\beta_R=0.001.$} \label{vksi_comb}
\end{figure}

We show, in figure \ref{vksi_comb}, the angular parts $<V(n_z)>$ of links in space-like directions versus $\xi$ for $\beta_h=0.28,$ that is in the broken phase. We depict three different curves, the uppermost corresponding to positions away from the monopole, specifically at $n_z=0,$ and the lowest one corresponding to positions near the monopole, the site with $n_z=10,$ while the parameter $B$ is set to the values $B=500.$ The intermediate curve corresponds to $B=100.$ Notice that, the sites away from the core (upper curve) lie in the broken phase for any value of $\xi.$ For small values of $\xi,$ say at $\xi\simeq 0.5,\ B=500,$ the sites near the monopole core lie in the symmetric phase, as indicated by the small values of $<V(n_z=10)>.$ However, if even smaller values of $\xi$ are considered, the values of $<V(n_z=10)>$ approach the ones for $n_z=0.$ This value of $\xi,$ where the minimum of the lower curve is situated, depends on the value of $B$ that has been used: in particular it is $\xi\simeq 1,$ when $B=500.$ For  larger values of $B$ we expect that the $\xi$ which yields the minimum $<V(n_z=10)>$ will move to smaller values. On the contrary, for smaller values of $B$ it will move to the right. This is exactly what the intermediate curve shows: the link angular variable  $<V(n_z)>$ does not take small values any more and the position of the minimum moves to the right. This observation may be understood, since the influence of the source depends on $\xi^2 B,$ as may be seen by inspection of the expression (\ref{act1}). This behaviour suggests that, below some value of $\xi,$ it becomes difficult to dynamically create a monopole, possibly because of its too large energy. If one insists in creating a dynamical monopole for a given small value of $\xi,$ one should use a sufficiently large $B.$ However, letting $B$ grow large means that the background will be too large near the ends of the lattice, unless one can use even bigger lattices. This is the reason why we have chosen to use consistently the values $\xi=1.00$ in our simulations and have not attempted to approach lower values of $\xi.$

The behaviour of the observables shows that the system lies in different phases at different regions of the lattice. This characteristic is less sharp for large values of $\xi,$ which drive the system to a definite phase throughout the lattice. One might also consider similar lines at distances between the ones depicted here $(0<n_z<10),$ and the result would be a set of curves filling the space between the curves of figure \ref{vksi_comb}. It is important to notice that there is a smooth transition between the regions with small and large values of $\xi.$ That means that the limit $\xi\to 0$ is more or less smooth and the characteristics of the model pertaining to $\xi=0$ appear already at small, but non-zero, values of $\xi.$


\section{An alternative model: Standard Kinetic coupling between the two U(1)${\rm s}$}\label{sec:FAFC}

In this section we consider for comparison an alternative model with kinetic mixing term of the type \eqref{ordinary} (in the continuum limit), with a fixed coefficient $\lambda/(q_e q_m)=1/\xi^2$ in our Lattice units, used so far.  There are no mixed CS terms of the form \eqref{CSterms} in this model. The action now reads: 
\begin{align}\label{FAFC} S & = \f{\beta_A}{2} \left(1 +\f{4}{\xi^2}\right) \sum_x \sum_{1\le\mu<\nu\le 4} [F_{\mu\nu}^{A,latt}(x)]^2 + \f{\beta_C}{2} \left(1 +\f{4}{\xi^2}\right) \sum_x \sum_{1\le\mu<\nu\le 4} [F_{\mu\nu}^{C,latt}(x)]^2 \nonumber \\ &+\f{2 \sqrt{\beta_A \beta_C}}{\xi^2} \sum_x \sum_{1\le\mu<\nu\le 4} [(F^{A,latt}_{\mu\nu}+ \chi F^{B,latt}_{\mu\nu}) F^{C,latt}_{\mu\nu}] \nonumber \\ &+ \sum_x \Phi^*(x)\Phi(x) - \beta_h \sum_x \left[\sum_{1\le \mu\le 4} \Phi^*(x) U^A_{x \hat{\mu}} \Phi(x+\hat{\mu})\right]  + \beta_R \sum_x [\Phi^*(x)\Phi(x)-1]^2.
\end{align} 

We start by studying the behaviour of the model without a monopole background as the parameter $\beta_h$ varies. We set $\beta_A=4.0,$ $\beta_C=0.25,$ $\beta_R=0.001,$ $B=0,$ and plot the quantities $<V^{(0)}>$ and $<R_2^{(0)}>$ defined in equations (\ref{V1}) and (\ref{R21}), versus $\beta_h$ for two values of $\xi,$ namely $\xi=1.00$ and $\xi=10.00.$ The plots show that the presence of the kinetic coupling does not influence the behaviour of the system much. In particular, one may be sure that for $\beta_h=0.28$ the system is in the broken phase.

\begin{figure}[ht]
\begin{center}
\includegraphics[scale=0.6,angle=0]{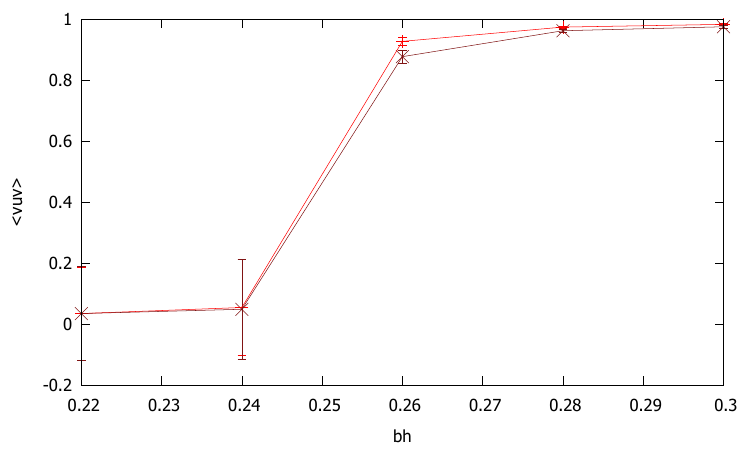}
\includegraphics[scale=0.6,angle=0]{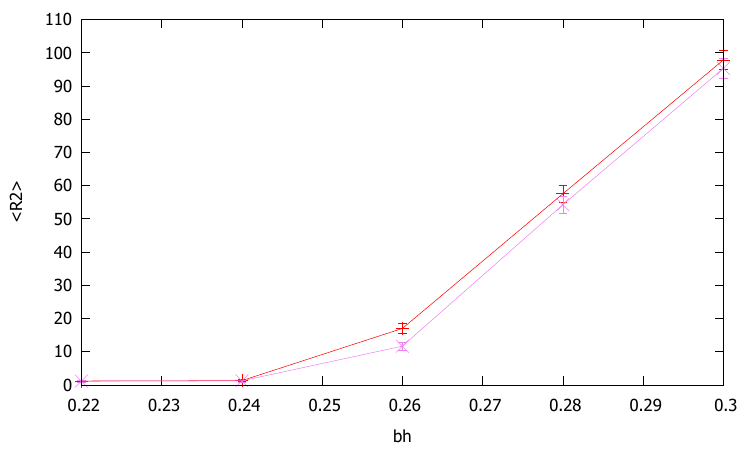}
\end{center}
\caption {Angular parts $<V^{(0)}>$ of links in space-like directions versus $\beta_h$ in the absence of the monopole background (left panel). Also $<R_2^{(0)}>$ are depicted (right panel). The parameters read: $\beta_A=4.0,$ $\beta_C=0.25,$ $B=0$ and $\beta_R=0.001.$ Two values for $\xi$ are depicted: $\xi=10.00$ (lower curves) and $\xi=1.00$ (upper curves).} \label{linksymmFF}
\end{figure}

\begin{figure}[ht]
\begin{center}
\includegraphics[scale=0.6,angle=0]{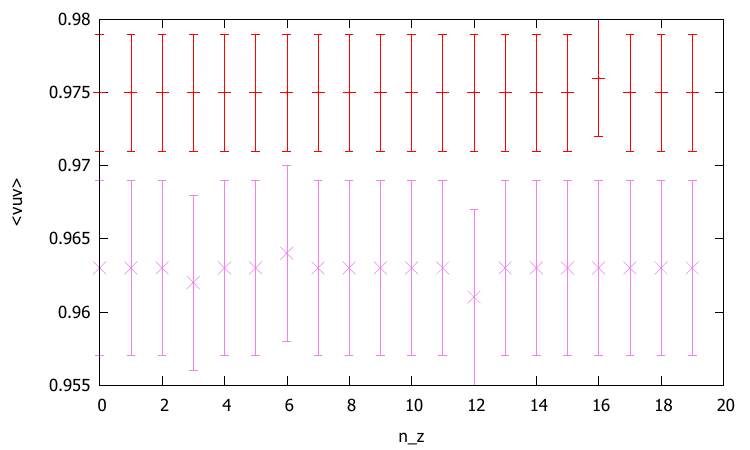}
\includegraphics[scale=0.6,angle=0]{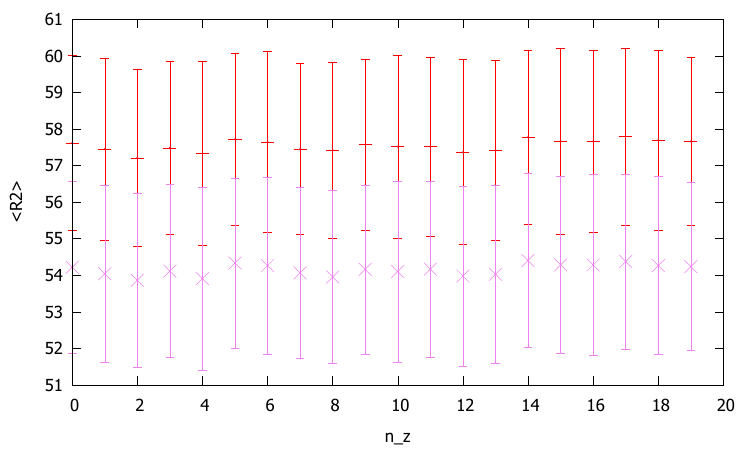}
\end{center}
\caption {Angular parts of links in space-like directions versus $n_z$ (left) in the broken phase $(\beta_h=0.28).$ Also $<R_2(n_z)>$ is depicted (right). The parameters read: $\beta_A=4.0,$ $\beta_C=0.25,$ $B=500$ and $\beta_R=0.001.$ The parameter $\xi$ takes the values $10.00,$ (the lowest curves) and $1.00$ (the uppermost curves).} \label{R2brFF}
\end{figure}

We now come to the investigation of the model with the background monopole source. We set $B=500$ and study the $n_z$ dependence of the quantities $<V(n_z)>$ and  $<R_2(n_z)>,$ defined via equations (\ref{Vz}) and (\ref{R2z}), for $\beta_A=4.0,$ $\beta_C=0.25,$ $\beta_h=0.28$ $\beta_R=0.001.$ Two values for $\xi$ have been considered, namely $\xi=1.00$ and $\xi=10.00.$ We have just considered the broken phase of the model, since in the symmetric phase nothing interesting happens, as we have checked, similarly to the previous model. Based on our previous experience we check whether there exists some dependence on $n_z$ of the angular parts of links in space-like directions and the measure squared of the scalar field. The difference between the two values of $\xi,$ shown in figure \ref{R2brFF} is just that the results for $\xi=1.00$ are larger than the corresponding results for $\xi=10.00,$ but no significant dependence on $n_z$ shows up, in contrast to the previous model. This is expected, since the kinetic coupling has an entirely different structure from the mixed CS coupling of the field strength of the potential $A$ with the dual field strength of the potential $C$, which characterised  the magnetic monopole case.

\section{Conclusions and Outlook}\label{sec:concl}

In this work, we have studied non perturbatively (on the lattice, {\it cf.} section \ref{sec:lattice}) a proposal for the description of the effects of a quantum fluctuating monopole on Higgs matter in a gauge theory of $U_{\rm em}(1) \times U_d(1)$, where the scalar Higgs field couples only to $U_{\rm em}(1)$ representing electromagnetism. The dual $U_d(1)$ has been argued to represent the quantum fluctuations of the magnetic monopole field, which is characterised by a non trivial background magnetic field of the characteristic singular type at the origin of the monopole core.

The use of two gauge potentials was inspired by the work of \cite{Zwanziger:1970hk}, in an attempt to avoid the use of Dirac strings. Following \cite{terning}, which argued on the appearance of the Lorentz-violating effects of the Dirac string only on the phase of the scattering amplitudes of a monopole off matter, or the absence of such effects altogether if the Dirac quantization condition \eqref{schquant} were in operation, we have considered the lattice version of the effective action \eqref{lag2} in an attempt to study the phase diagram of charged scalar matter interacting with a magnetic monopole. 
The latter was described both, by an external background magnetic field, with the characteristic monopole singular structure at the origin, \eqref{monmagn}, and by quantum fluctuations described by the gauge potential $\mathcal C^\mu$ of the $U_d(1)$ gauge group. 

It is important to notice that in our approach, which generalises non trivially the original work of \cite{Zwanziger:1970hk}, the constraint \eqref{solconstr} (upon ignoring the Dirac-string effects) among the two gauge potentials is implemented in a path integral via the introduction of the gauge- and Lorentz-invariant $\delta$-functional term \eqref{deltarep}. However, in our generalised analysis we consider the parameter $\xi$ as taking values in the entire real axis, not only in the region $\xi^2 \to 0$ that defines the monopole case of  \cite{Zwanziger:1970hk}. Such an effective description ({\it cf.} \eqref{lag2}) implies the existence of CS terms mixing the electromagnetic with the dual gauge field strengths. As discussed in section \ref{sec:FAFC}, such terms are important in yielding configurations in the matter field, with a behaviour representing the emergence of a magnetic monopole configuration.  The appearance of such configurations are {\it triggered} in our approach by the monopole background source \eqref{monmagn}. The situation is similar to the microscopic monopoles, of e.g. t'Hooft-Polyakov type~\cite{thooft} in models with adjoint Higgs fields, which are solutions of the classical equations. In the lattice version of such models, the `t Hooft-Polyakov monopole configurations would appear upon triggering with appropriate external sources. 

In our case we treat the background monopole source as a Dirac, point-like one, without specifying any structure. In the case of `t Hooft-Polyakov models, e.g. in the SU(2) case with scalar triplets, the latter lead to the well known homotopy $\Pi_2 (SU(2))$ properties leading to the magnetic monopole sectors of the non-trivial solutions. In our Dirac-source case, we are agnostic as to the precise microscopic homotopy structure of the monopole configuration arising in our Lattice simulation in the broken phase ({\it cf.} 
section \ref{sec:higgs}, in particular subsection \ref{sec:broken}). In subsection \ref{sec:xidep}, we have seen that the emergence of a non-trivial configuration for the scalar field, with a behaviour familiar from the t' Hooft-Polyakov-monopole case, is triggered by relatively strong source magnetic 
fields, e.g. for $\xi=1$ we need a magnetic intensity of the source of order $B={500}$ in lattice units ({\it cf.} \eqref{maglatt}).
Deep in the broken phase, or equivalently for smaller $\xi$ ({\it cf.} \ref{sec:broken}), one needs much stronger sources to be able to see monopole configurations. Because the magnetic field carries energy of order $B^2$ one expects from such arguments that the 
magnetic monopole configuration in our case is sufficiently massive in appropriate units.

\section*{Acknowledgements}

 The work of G.K. is supported by the research project of the National Technical University of Athens (NTUA)
65232600-ACT-MTG, entitled: Alleviating Cosmological Tensions Through Modified Theories of Gravity.
The work of N.E.M. is supported in part by the UK Science and Technology Facilities research
Council (STFC) and UK Engineering and Physical Sciences Research
Council (EPSRC) under the research grants ST/X000753/1 and  EP/V002821/1, respectively.

\appendix 

\section{Simulation details}\label{sec:simul}

The gauge part of the action reads: 
\begin{align} S_g &= \f{\beta_A}{2} \left(1 +\f{4}{\xi^2}\right) \sum_x \sum_{1\le\mu<\nu\le 4} [F_{\mu\nu}^{A,latt}(x)]^2 + \f{\beta_C}{2} \left(1 +\f{4}{\xi^2}\right) \sum_x \sum_{1\le\mu<\nu\le 4} [F_{\mu\nu}^{C,latt}(x)]^2\nonumber \\&+\f{2\, \sqrt{\beta_A\, \beta_C}}{\xi^2} \sum_x \sum_{1\le\mu<\nu\le 4} \sum_{\rho,\sigma} \epsilon_{\mu\nu\rho\sigma} [(F^{A,latt}_{\mu\nu}(x)+ \chi F^{B,latt}_{\mu\nu}(x)) F^{C,latt}_{\rho\sigma}(x)],\end{align} where $F_{\mu\nu}^{A,latt}(x),$ $F_{\mu\nu}^{B,latt}(x)$ and $F_{\mu\nu}^{C,latt}(x)$ represent the lattice versions of the field strengths, and we remind the reader that $\chi=\xi^2/4$. In the following we denote them simply by $F_{\mu\nu}^{A}(x),$ $F_{\mu\nu}^{B}(x)$ and $F^C_{\mu\nu}$. 

We concentrate on the last part of the action. We want to simulate the mixed CS term $$ \f{2}{\xi^2} \sum_{\mu,\nu} (F^A_{\mu\nu}+ \chi F^B_{\mu\nu})\cdot(\tilde{F}^{C}_{\mu\nu}) = \f{2}{\xi^2} \sum_{\mu<\nu} \epsilon_{\mu\nu\rho\sigma} (F^A_{\mu\nu}+\chi F^B_{\mu\nu}) F^C_{\rho\sigma},$$ since $\tilde{F}^{C}_{\mu\nu} = \f{1}{2} \sum_{\rho,\sigma} \epsilon_{\mu\nu\rho\sigma} F^C_{\rho\sigma}.$ 

We observe that, if we interchange e.g. $\mu$ and $\nu,$ the part of the action $\epsilon_{\nu\mu\rho\sigma} F^A_{\nu\mu} F^C_{\rho\sigma}$ is equal to $\epsilon_{\mu\nu\rho\sigma} F^A_{\mu\nu} F^C_{\rho\sigma},$ because of the antisymmetry of both $\epsilon_{\mu\nu\rho\sigma}$ and $F^A_{\mu\nu}.$ With similar arguments we conclude that the (relevant part of the) action equals: 
\begin{align} S_{{\rm mixed}~CS} &= \f{2}{\xi^2} [(F^A_{12}+ \chi F^B_{12}) F^C_{34} + (F^A_{13}+ \chi F^B_{13}) F^C_{42} + (F^A_{14} + \chi F^B_{14}) F^C_{23} \nonumber \\ &+ (F^A_{34}+ \chi F^B_{34}) F^C_{12} + (F^A_{42}+ \chi F^B_{42}) F^C_{13} + (F^A_{23} + \chi F^B_{23}) F^C_{14}].\label{action}\end{align}

Let us consider updating of the A field. We shall change in turn the links $A^A_1(x),\ A^A_2(x),\ A^A_3(x)$ and $A^A_4(x).$ We recall the definition: \be F^A_{\mu\nu} = A^A_\nu(x+\hat{\mu}) -A^A_\nu(x) - A^A_\mu(x+\hat{\nu}) + A^A_\mu(x).\ee We observe that, upon the replacement: $$ A^A_\mu(x)\to A^A_\mu(x) + \Delta\Rightarrow \delta A^A_\mu(x) = \Delta,$$ i.e. the field strength  $F^A_{\mu\nu}$ will change by $+\Delta.$ If we replace $$ A^A_\nu(x)\to A^A_\nu(x) + \Delta\Rightarrow \delta A^A_\nu(x) = \Delta,$$ the quantity $F^A_{\mu\nu}$ will change by $-\Delta.$

In general the replacement $ \delta A^A_\mu(x) = + \Delta$ implies $ \delta F^A_{\mu\nu}(x) = +  \Delta,$ while the replacement $ \delta A^A_\nu(x) = + \Delta$ implies $ \delta F^A_{\mu\nu}(x) = - \Delta.$

As stated above, for each $x,$ we will update successively the links $A^A_1(x),\ A^A_2(x),\ A^A_3(x)$ and $A^A_4(x).$

\begin{itemize} 
\item{\bf (a)} When we update $A^A_1(x),$ some terms in (\ref{action}) that will be influenced will be: $F^A_{12}(x) F^C_{34}(x),$ $F^A_{13}(x) F^C_{42}(x)$ and $F^A_{14}(x) F^C_{23}(x).$ The change in the action due to these terms only will be: $$ \f{2}{\xi^2} ( \delta F^A_{12}(x) F^C_{34}(x) + \delta F^A_{13}(x) F^C_{42}(x) + \delta F^A_{14}(x) F^C_{23}(x)) = \f{2}{\xi^2} \Delta (F^C_{34}(x) + F^C_{42}(x) + F^C_{23}(x)).$$

In addition the terms e.g. $F^A_{12}(x-\hat{2}) F^C_{34}(x-\hat{2})$ contain $A^A_1(x),$ but the sign of the corresponding change will be opposite: $-\Delta F^C_{34}(x-\hat{2}),$ so that this kind of terms yield: $\Delta [F^C_{34}(x) - F^C_{34}(x-\hat{2})].$ 

The terms $F^A_{13}(x) F^C_{42}(x)$ will yield additional contributions, so that this kind of terms yield: $\Delta [F^C_{42}(x) - F^C_{42}(x-\hat{3})].$  Similarly the terms $\Delta F^C_{23}(x)$ should be completed to: $\Delta [F^C_{23}(x) - F^C_{23}(x-\hat{4})]$

Finally the change in the action reads: $$ \delta S_1^A = \f{2}{\xi^2} \Delta \left\{ [F^C_{34}(x) - F^C_{34}(x-\hat{2})] + [F^C_{42}(x)-F^C_{42}(x-\hat{3})] + [F^C_{23}(x) - F^C_{23}(x-\hat{4})]\right\}.$$

\item{\bf (b)} When we update $A^A_2(x),$ some terms in (\ref{action}) that will change will be $F^A_{12} F^C_{34},\ F^A_{42} F^C_{13}$ and $F^A_{23} F^C_{14}.$ The change in the action will be $$ \f{2}{\xi^2} \Delta (-F^C_{34} - F^C_{13} + F^C_{14}).$$ The signs are reflections of the result that, in this case: $\delta F^A_{12}=-\Delta,$ $\delta F^A_{42}=-\Delta$ and $\delta F^A_{23}=+\Delta.$

As before, there is more, so that: 
$$\delta S_2^A = \f{2}{\xi^2} \Delta \left\{ -[F^C_{34}(x)-F^C_{34}(x-\hat{1})] - [F^C_{13}(x)-F^C_{13}(x-\hat{4})] + [F^C_{14}(x)-F^C_{14}(x-\hat{3})]\right\}.$$

\item{\bf (c)} When we update $A^A_3(x),$ the terms in (\ref{action}) that will change will be $F^A_{13} F^C_{42},\ F^A_{34} F^C_{12}$ and $F^A_{23} F^C_{14}.$ The change in the action will be $$ \f{2}{\xi^2} \Delta (-F^C_{42} + F^C_{12} - F^C_{14}).$$ The signs are reflections of the result that, in this case: $\delta F^A_{13}=-\Delta,$ $\delta F^A_{34}=+\Delta$ and $\delta F^A_{23}=-\Delta.$ 

Including the additional contributions we end up with: 
$$\delta S_3^A = \f{2}{\xi^2} \Delta \left\{ -[F^C_{42}(x)-F^C_{42}(x-\hat{1})] + [F^C_{12}(x)-F^C_{12}(x-\hat{4})] - [F^C_{14}(x)-F^C_{14}(x-\hat{2})]\right\}.$$

\item{\bf (d)} When we update $A^A_4(x),$ the terms in (\ref{action}) that will change will be $F^A_{14} F^C_{23},\ F^A_{34} F^C_{12}$ and $F^A_{42} F^C_{13}.$ The change in the action will be $$ \f{2}{\xi^2} \Delta (-F^C_{23} - F^C_{12} + F^C_{13}).$$ The signs are reflections of the result that, in this case: $\delta F^A_{14}=-\Delta,$ $\delta F^A_{34}=-\Delta$ and $\delta F^A_{42}=+\Delta.$ 

Including the additional contributions we end up with:
$$\delta S_4^A = \f{2}{\xi^2} \Delta \left\{ -[F^C_{23}(x)-F^C_{23}(x-\hat{1})] - [F^C_{12}(x)-F^C_{12}(x-\hat{3})] + [F^C_{13}(x)-F^C_{13}(x-\hat{2})]\right\}.$$

We now proceed with the updating of the $C$ field. We will update successively the links $A^C_1(x),\ A^C_2(x),\ A^C_3(x)$ and $A^C_4(x).$

\item{\bf (e)}$$ \delta S_1^C = \f{2}{\xi^2} \Delta \left\{ [(F^A_{34}(x)+ \chi F^B_{34}(x)) - (F^A_{34}(x-\hat{2})+ \chi F^B_{34}(x-\hat{2}))] \right.$$ 
$$ + [(F^A_{42}(x) + \chi F^B_{42}(x))-(F^A_{42}(x-\hat{3})+ \chi F^B_{42}(x-\hat{3}))] $$ $$ \left. + [(F^A_{23}(x)+ \chi F^B_{23}(x)) - (F^A_{23}(x-\hat{4})+ \chi F^B_{23}(x-\hat{4}))]\right\}.$$

\item{\bf (f)} $$\delta S_2^C = \f{2}{\xi^2} \Delta \left\{ -[(F^A_{34}(x)+ \chi F^B_{34}(x))-(F^A_{34}(x-\hat{1})+ \chi F^B_{34}(x-\hat{1}))]\right.$$ $$ - [(F^A_{13}(x)+ \chi F^B_{13}(x))-(F^A_{13}(x-\hat{4})+ \chi F^B_{13}(x-\hat{4}))] $$ $$ \left.+ [(F^A_{14}(x)+\chi F^B_{14}(x))-(F^A_{14}(x-\hat{3}) + \chi F^B_{14}(x-\hat{3}))]\right\}.$$

\item{\bf (g)} $$\delta S_3^C = \f{2}{\xi^2} \Delta \left\{ -[(F^A_{42}(x)+ \chi F^B_{42}(x))-(F^A_{42}(x-\hat{1})+\chi F^B_{42}(x-\hat{1}))]\right.$$ $$ + [(F^A_{12}(x)+\chi F^B_{12}(x))-(F^A_{12}(x-\hat{4})+ \chi F^B_{12}(x-\hat{4}))]$$ $$\left. - [(F^A_{14}(x)+ \chi F^B_{14}(x))-(F^A_{14}(x-\hat{2})+ \chi F^B_{14}(x-\hat{2}))]\right\}.$$

\item{\bf (h)} $$\delta S_4^C = \f{2}{\xi^2} \Delta \left\{ -[(F^A_{23}(x)+ \chi F^B_{23}(x))-(F^A_{23}(x-\hat{1})+ \chi F^B_{23}(x-\hat{1}))]\right.$$ $$ - [(F^A_{12}(x)+ \chi F^B_{12}(x))-(F^A_{12}(x-\hat{3})+ \chi F^B_{12}(x-\hat{3}))]$$ $$\left. + [(F^A_{13}(x)+ \chi F^B_{13}(x))-(F^A_{13}(x-\hat{2})+\chi F^B_{13}(x-\hat{2}))]\right\}.$$

\end{itemize}

\end{document}